\documentclass{aa}
\usepackage{graphicx} % Include figure files
\usepackage{dcolumn}  % Align table columns on decimal point
\usepackage{bm}       % bold math
\usepackage{amssymb,amsmath}
\usepackage{natbib}
\newcommand{\ontop}[2]{
  \renewcommand{\arraystretch}{0.2}
  \begin{array}{c}
  #1 \\ #2
  \end{array}
  \renewcommand{\arraystretch}{1.0}
}
\newcommand{\lsim}{\ontop{<}{\sim}}
\newcommand{\mat}[1]{\bm{\mathrm{#1}}}

\def\galsim{\textsc{Galsim}}

\def\ksb{\textsc{KSB}}

\def\shapelens{\textsc{shapelens}}

\begin{document}

\title{A highly precise shear bias estimator independent of the measured shape noise}
\titlerunning{Shear bias estimator}

\authorrunning
{Arnau Pujol\and
Martin Kilbinger\and
Florent Sureau\and
et al.
}
\author
{Arnau Pujol \inst{1,2} \and
Martin Kilbinger \inst{1,2,3} \and
Florent Sureau \inst{1,2} \and
Jerome Bobin \inst{1,2}  \\
}
\institute{
DEDIP/DAP, IRFU, CEA, Universit\'e Paris-Saclay, F-91191 Gif-sur- Yvette, France\\
\and
Universit\'e Paris Diderot, AIM, Sorbonne Paris Cit\'e, CEA, CNRS, F-91191 Gif-sur-Yvette, France\\
\and
Institut d'Astrophysique de Paris, UMR7095 CNRS, Universit\'e Pierre \& Marie Curie, 98 bis boulevard Arago, F-75014 Paris, France\\
}

\date{Received date / Accepted date}

\abstract{
We present a new method to estimate shear measurement bias in image
simulations that significantly improves the precision with respect to
current techniques. Our method is based on measuring the shear response
for individual images. We generated sheared versions of the same image to
measure how the galaxy shape changes with the small applied shear. This
shear response is the multiplicative shear bias for each image. In addition, we also measured
the individual additive bias. Using the  same noise realizations for each sheared version
allows us to compute the shear response at very high precision. The estimated
shear bias of a sample of galaxies is then the average of the
individual measurements. The precision of this method
leads to an improvement with respect to previous methods concerned with the precision of estimates of multiplicative bias since our method is
not affected by noise from shape measurements, which until now has been the dominant uncertainty.
As a consequence, the method does not require
shape-noise suppression for a precise estimation of shear multiplicative bias. Our method can
be readily used for numerous applications such as shear measurement validation and
calibration, reducing the number of necessary simulated images by a few
orders of magnitude to achieve the same precision.
}

\keywords{
weak gravitational lensing - shear bias}

\maketitle

\section{Introduction}

Upcoming weak-lensing surveys have the goal of measuring cosmology with
unprecedentedly high precision. Their very high statistical power requires
systematic errors to be very well understood and calibrated. One of the main
sources of systematic error for weak gravitational lensing is the bias in the
measurement of the galaxy shear, which carries the cosmological information
about the galaxy's large-scale structure and its evolution. For upcoming experiments
such as Euclid \citep{Laureijs2011}, the Large Synoptic Survey Telescope \citep[LSST,][]{LSST2009}, or
the Wide Field Infrared Survey Telescope  \citep[WFIRST,][]{2013arXiv1305.5422S}, we need to calibrate shear biases to
sub-percent precision. Traditional calibration methods create large suites of
galaxy image simulations and estimate the shear bias for a given shape
measurement method, point spread function (PSF), galaxy population, noise level, and other factors. Shear bias
estimation to date has been dominated by the intrinsic ellipticity dispersion of the
simulated galaxies. To reach the desired precision requires a simulation volume
that exceeds the actual observational data that are to be calibrated, with
billions of simulated galaxies. This has a dramatic impact on the computation
load of generating both the simulations and shape measurement methods,
and therefore on limiting the complexity, storage, and re-usability
\citep[e.g.][]{Hoekstra2017}.

Existing methods to reduce the number of simulations made use of rotated
galaxies with the same shear such that their mean intrinsic ellipticity cancels out differences. The first proposed method was the so-called ring test
\citep{2007AJ....133.1763N}, using galaxies evenly distributed in their
orientation of intrinsic ellipticity with constant modulus, and with constant
shear. \citet{Massey2007b} reduced the number of objects to a pair of
orthogonally oriented galaxies. Both approaches result in a zero net intrinsic
ellipticity. This reduces the error of the estimated shear, but does not
entirely cancel out the contribution from the measured shapes.
Stochasticity in the measurement, for example~due to pixel noise and the PSF, will
perturb the exact shape-noise cancellation. In addition, systematic biases
break the input ellipticity symmetry. Sources of these systematic effects are
ellipticity bias (the response of the measurement to intrinsic ellipticity)
when it depends on the orientation of the galaxy with respect to the coordinate
system, the PSF, or the shear, and also selection effects, including
non-equal galaxy weights. The selection-induced shear bias calibration
for the Hype Suprime-Cam (HSC) survey was performed without shape-noise suppression \citep{2017arXiv171000885M}.

In the case of simulating fields with non-constant shear, noise suppression can be
achieved by simulating the intrinsic shape-noise distribution of galaxies as a
pure B-mode field. Using an estimate of the E-mode power spectrum or real-space
E-mode correlation is then insensitive to the intrinsic shape noise
\citep{Kitching2010}. However, simulating a realistic intrinsic ellipticity
distribution inevitably leads to a power leakage from B to E
\citep{Mandelbaum2014}. In addition, measurement stochasticity and biases,
causing imperfect noise cancellation as described above, also apply in the case
of variable shear.

In this paper we propose a new method to estimate the shear bias from
simulations that is insensitive to the shear estimator noise coming from both the
intrinsic and the measured ellipticity dispersion. This reduces the number of
required simulated galaxies by three orders of magnitude. Even though we
estimate a bias for each galaxy individually, for calibration purposes we only
use the mean bias averaged over a sample of galaxies. This avoids unstable
ratios of two noisy quantities.

Our method is inspired by the metacalibration technique
\citep{Sheldon2017,Huff2017}, where the shear bias is estimated as the shape
estimator response to a small shear applied directly to the individual images. This technique is used to calibrate shear bias on real data without the need to create simulations \citep{Zuntz2017}. However, it requires us to perform operations on the images such as PSF de- and re-convolution, and subtraction and addition of noise components.
Our method is not a calibration technique, but a shear bias estimation. As it is simulation-based, we do not need to de-convolve
and de-noise observed images with an estimated PSF, which is notoriously
difficult. We can apply any shear to the simulated image before the PSF
convolution step. The challenge for shear calibration using our method, as with all simulation-based techniques, is to closely match the properties
of the simulations to the observed data in order to minimize the important biases due to selection effect (see e.g.~\cite{Conti2016}).

This paper is organized as follows. In Sect.~\ref{sec:defs} we define the basic
required concepts. Section~\ref{sec:shear_bias_measurement} introduces our new
shear bias estimation method and contrasts it with existing ones. In
Sect.~\ref{sec:error_estimation} we analytically compute the precision of our
method and compare it to existing shear bias estimation techniques. In
Sect.~\ref{sec:data} we describe the galaxy image simulations, which we use in
Sect.~\ref{sec:results} to test our analytical descriptions, and to compare our
method with existing ones. We discuss potential applications of our method in
Sect.~\ref{sec:applications}, and give a summary in
Sect.~\ref{sec:conclusions}.
\section{Definitions}\label{sec:defs}

\subsection{Shear bias}

We define multiplicative and additive shear bias for a population of galaxies.
Let $g_\alpha$ be the shear of a given galaxy and $e_{\alpha}^{\rm obs}$ its
observed ellipticity, where $\alpha = {1,2}$ stands for the two components of
the complex shear and ellipticity. If the mean intrinsic ellipticity $\langle
e_\alpha^{\rm I} \rangle$ of the galaxy sample is zero, we can estimate the
mean reduced shear $g_\alpha^{\rm obs}$ as the average of the observed
ellipticities\footnote{The following commonly used equation ignores the $2\times2$-tensor nature of $m$.
We will use the full expression for the shear response defined below.},
\begin{equation}
g_{\alpha}^{\rm obs} \equiv \langle e_{\alpha}^{\rm obs} \rangle =  c_\alpha + (1 + m_\alpha) \langle g_\alpha \rangle .
\label{eq:g_relation}
\end{equation}
This estimator is biased in general, and $c_{\alpha}, m_{\alpha}$ are the
ensemble-average additive and multiplicative shear biases
\citep{Huterer2006,Heymans2006}. Here and in the following, we ignore
non-linear contributions to the bias. We also neglect higher-order terms,
for example~the term $\bm g^\ast \bm e^{\rm I}$ in the denominator of the
shear estimator introduced in \citet{1997A&A...318..687S}. Further we set the
convergence $\kappa = 0$ such that the (observable) reduced shear $\bm g = \bm
\gamma / (1 - \kappa)$ equals the shear $\bm \gamma$.

Alternatively, we can describe a shear bias for individual galaxies as the
response of the observed ellipticity to a small shear distortion
\citep{Huff2017,Sheldon2017}:
\begin{equation}
R_{\alpha \beta}  =  \frac{\partial e_\alpha^{\rm obs}}{\partial g_\beta}.
\label{eq:ind_bias}
\end{equation}
The shear response $\mat R$ is a $2\times2$ matrix whose diagonal (off-diagonal) terms represent
the response of the ellipticity measurements to shear changes of the same
(opposite) component.
An additive shear bias for individual galaxies is defined as
%
%\begin{equation}
%a_\alpha = e^{\rm obs}_\alpha - e^{\rm I}_\alpha,
%\label{sec:additive}
%\end{equation}
%
%when $\bm g = 0$, where $e^{\rm I}_\alpha$ is the intrinsic ellipticity. Alternatively, it can be obtained from
%
\begin{equation}
a_\alpha = e^{\rm obs}_\alpha - R_{\alpha \alpha}g_\alpha - e^{\rm I}_\alpha,
\label{sec:additive}
\end{equation}
where $e^{\rm I}_\alpha$ is the intrinsic ellipticity.
%when $\bm g \neq 0$.
We note that $e^{\rm I}_\alpha$ can be defined if we use an analytic expression for the galaxy profile, but complex galaxy morphologies do not necessarily have a unique `true' ellipticity, in which case we cannot measure $a_\alpha$ using Eq. 3.
We can however estimate the additive bias if
$\langle e^{\rm I}_\alpha \rangle = 0$ over the population, which can be fulfilled under certain symmetry assumptions without the need to define intrinsic ellipticity, as we describe later in Sect. \ref{sec:our_method}.

A perfect shape estimation corresponds to $\mat R$ being the unit matrix and $a_\alpha = 0$.
If the shape measurement conserves the spin-2 property of
ellipticity and shear, $R_{\alpha \beta}$ needs to be a combination of a scalar and a
spin-4 tensor. If we neglect the latter, the response collapses to a single
non-zero number $R_{11} = R_{22}$, with $R_{12} = R_{21} = 0$.

\subsection{Shear calibration}

A simulation-based calibration of measured shear estimates typically measures
ensemble biases $m_\alpha , c_\alpha$ from a large
number of image simulations with different galaxy properties and shear $g_\alpha$
via Eq. 1. A calibrated shear estimate is then obtained by
correcting the observed ellipticities by the ensemble biases,
$(e_\alpha^{\rm obs} -  c_\alpha  ) / (1 +  m_\alpha )$, provided
the simulated galaxy population matches the data in all relevant properties.

When measuring the individual responses using Eqs. \ref{eq:ind_bias} and
\ref{sec:additive}, an unbiased shear estimator is given as $\langle \mat R
\rangle^{-1} \langle \bm e^{\rm obs} - \langle \bm a \rangle \rangle \approx
\langle \mat R \rangle^{-1} \langle \mat R \bm \gamma \rangle$ (see
\cite{Sheldon2017}). Below we present our method to compute $\mat R$ for each
simulated galaxy without being sensitive to shape noise.

\section{Shear bias measurement methods}\label{sec:shear_bias_measurement}

\subsection{Our method: Shear bias estimation reducing measurement noise}\label{sec:our_method}

We measure the shear response using Eq. 2 for individual simulated
galaxy images as follows. For each simulated galaxy with given properties,
intrinsic ellipticity $\bm e^{\rm I}$ and given shear $\bm g$ (which can
but need not be zero), we create additional, sheared versions of the same
galaxy. The galaxy is analytically sheared before PSF convolution and noise
addition, so that the differences between the images  only come from the
shear. We then approximate the shear response by finite
differences, following \cite{Huff2017},

\begin{equation}
  R_{\alpha\beta} \approx \frac{e^{\rm obs, +}_\alpha - e^{\rm obs, -}_\alpha}{2 \Delta g_\beta},
  \label{eq:ind_bias_estim}
\end{equation}
where $e^{\rm obs, \pm}_\alpha$ is the measured ellipticity of the image with
additional small shear $\pm \Delta g_\alpha$.
With three sheared images we can estimate all components of $\mat R$ for each galaxy. To determine the shear response
averaged over a sample of galaxies, we only require two appropriately chosen shear values
(see Sect.~\ref{sec:data} and Appendix~\ref{sec:robustness_m1} for more details).

To further reduce the stochasticity of our response estimator, we use the same
noise realization for all image copies for each galaxy. This guarantees that the contribution from the intrinsic ellipticity
exactly cancels out for  our bias estimator. Then, the intrinsic
ellipticity can be considered as just another property of the galaxy (like
the flux or radius) and as such affects the shear bias in a deterministic
way, but does not contribute to the statistical uncertainty. Therefore, we can
obtain a much more precise bias estimation compared to methods that average
over observed galaxy ellipticities. This will be quantified in
Sect.~\ref{sec:error_estimation}.

When randomizing the noise for each image, we obtain the same mean but noisier
response values. Keeping the same noise realization of our images is not
an artificial noise reduction in the bias estimate, it only helps us to obtain
a noise-free numerical derivative. The noise properties will be sufficiently
well sampled by the different simulated galaxies.
The additive shear bias for each galaxy is measured using Eq. 3
on the original, non-sheared image.

\begin{figure}
\centering
\includegraphics[width=.98\linewidth]{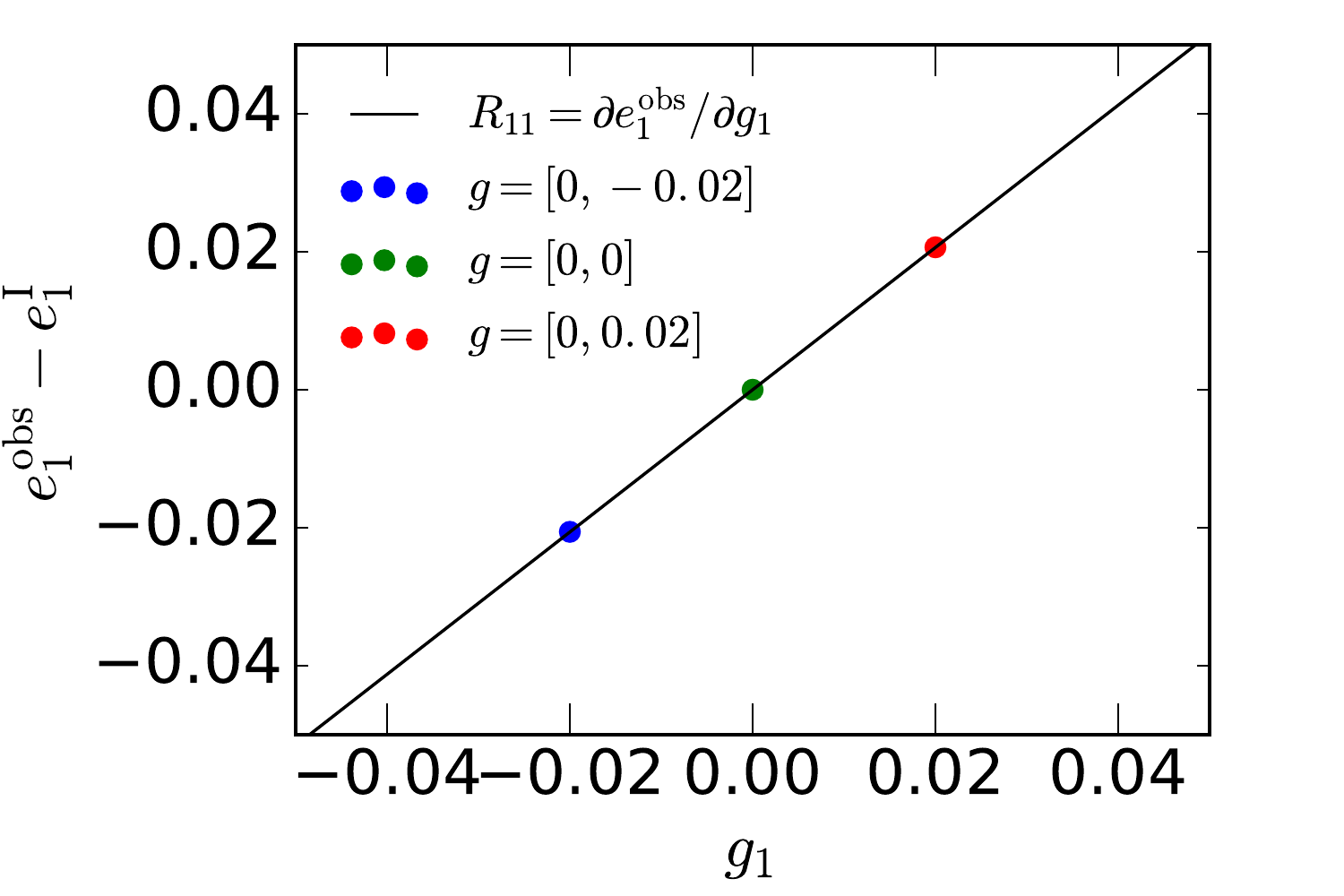}
\caption{Scheme of shear response estimation for a single galaxy  for $R_{11}$. }
\label{fig:b_scheme}
\end{figure}

In Fig. \ref{fig:b_scheme} we show an example of the estimated component of the
response, $R_{11}$, for one galaxy image. The finite-difference
estimate is insensitive to the shear value as long as it is small, $|\Delta
g_\alpha| \lsim 0.05$ for $\alpha=1, 2$. More details about the robustness of
our new estimator are presented in Appendix~\ref{sec:robustness_m1}.

From the measurements of individual galaxy shear biases, we estimate the
ensemble multiplicative and additive bias of a galaxy population as the average
of the individual estimates, respectively $\langle R_{\alpha\alpha} \rangle$ and $\langle a_\alpha
\rangle$. As mentioned before, we do not need to define $e^{\rm I}_\alpha$ as far as $\langle e^{\rm I}_\alpha \rangle = 0$ since then $\langle a_\alpha \rangle = \langle e^{\rm obs}_\alpha \rangle$. This is true for the usual cases of study with randomly oriented galaxies (assuming $\bm e^{\rm I}$ transforms under rotations like a spin-2 quantity) when the shape estimators have no preferred direction (which is something expected for most of the estimators).
This can be a weighted average if galaxies have different weights.
We ignore the non-diagonal terms of $\mat R$, as we have found that their
contribution averages out to zero if the shear values are
symmetrical around zero (see Appendix~\ref{sec:robustness_m1}).
In the following two subsections, we review two commonly used calibration methods
to estimate the shear bias.

\subsection{Linear fit estimation}
\label{sec:linear_fit}

\begin{figure}
\centering
\includegraphics[width=.98\linewidth]{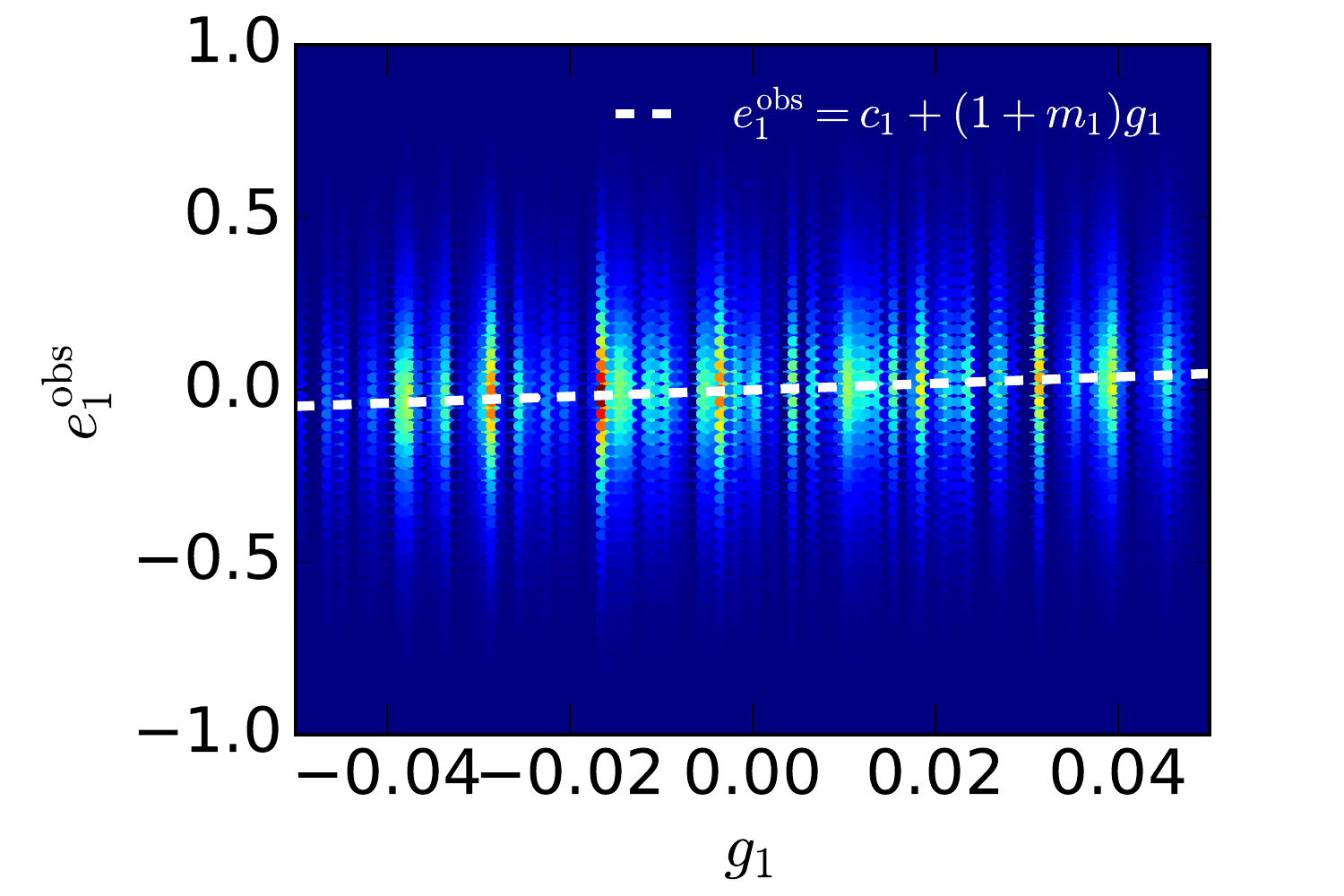}
\includegraphics[width=.98\linewidth]{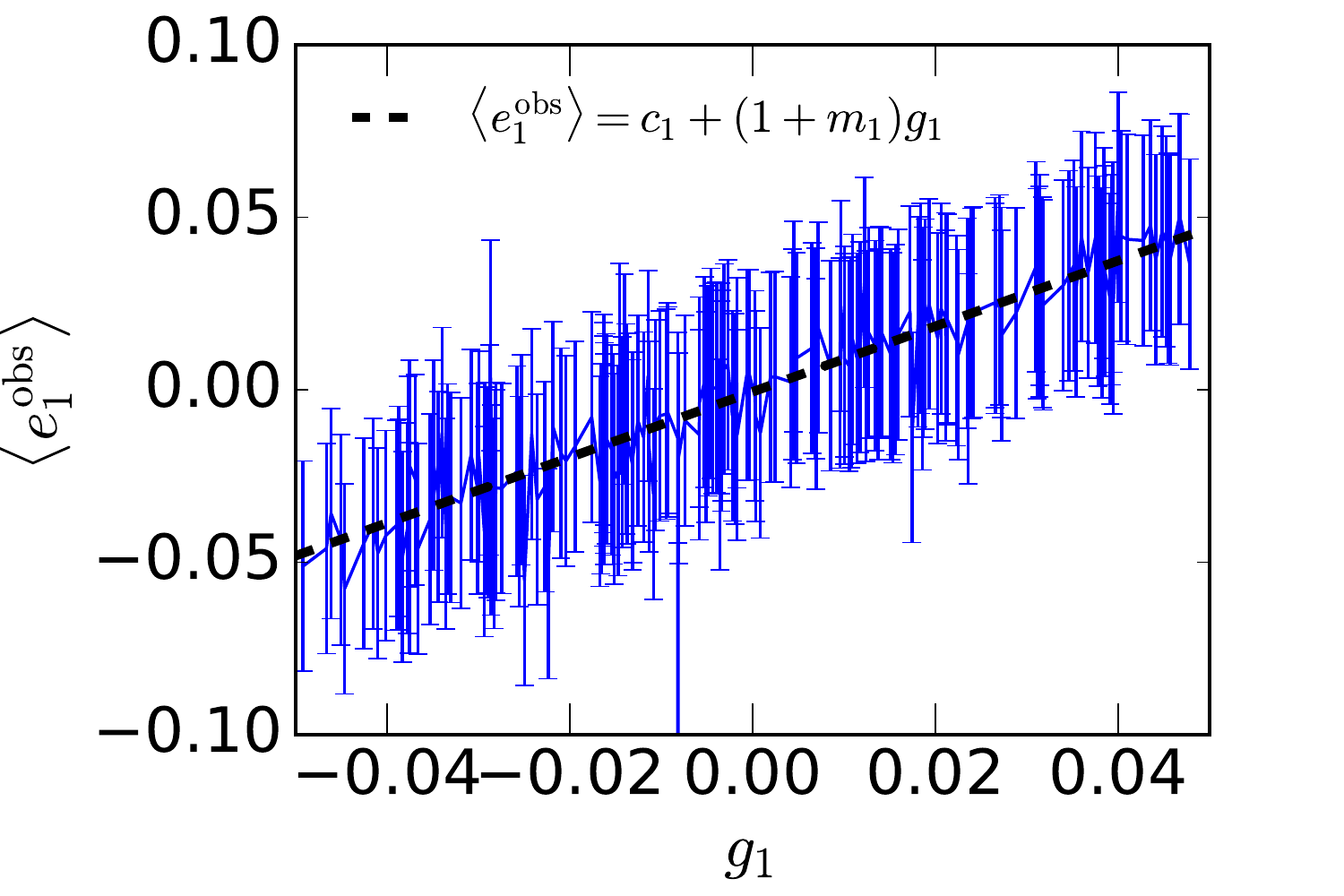}
\caption{Scheme of the estimation of biases $m_1$ and $c_1$ from the linear fit
 of the distribution of $e^{\rm obs}_1$ as a
function of $g_1$. In the top panel, the white dashed line shows the linear fit of the distribution,  represented by the colour map. In the bottom panel, the blue points and error bars show the mean $\langle e^{\rm obs}_1\rangle$ of the galaxies with the same shear. The black dashed line shows the weighted linear fit of the blue points. }

\label{fig:m2_scheme}
\end{figure}

The most common method to estimate the shear bias in the literature is to
perform a linear fit of Eq. 1 to simulated sheared galaxy
images
\citep[e.g.][]{Heymans2006,CFHTLenS-shapes,Zuntz2013,Mandelbaum2015,Conti2016,Huff2017,Hoekstra2017,Pujol2017,Zuntz2017,2017arXiv171000885M}.
For each galaxy population (e.g.~for each bin of given galaxy properties) we
obtain the additive and multiplicative biases $c_\alpha$ and $m_\alpha$ from a       linear
fit of the measured ellipticities as a function of simulated input shear, as
illustrated in the top panel of Fig.~\ref{fig:m2_scheme}. The error of the
parameter estimation can then be obtained by jackknife resampling and
obtaining the distribution of
best-fit parameters for each resample.

Alternatively, the straight line can be fitted to the average measured
ellipticities for each input shear, $\langle e^{\rm obs}_\alpha \rangle$, as
shown in the bottom panel of Fig.~\ref{fig:m2_scheme}. Both fitting schemes
provide consistent values and error bars for the shear bias parameters.

\subsection{Linear fit estimation with shape-noise suppression}
\label{sec:lin_fit_orth}

The precision of the linear fitting technique to measure shear bias is limited
by shape noise stemming from the intrinsic ellipticity distribution. Reducing this noise requires the use of a very large number of galaxy images. An
alternative method to reduce the shape-noise contribution is to force the mean
ellipticity to cancel out, by simulating orthogonal pairs of galaxy images
\citep{Massey2007b,Mandelbaum2014}, As described in \cite{Massey2007b}, the
estimated shear of a pair of orthogonal objects is
\begin{equation}
g^{\rm obs}_\alpha = \frac{e^{\rm obs}_{\alpha, A} + e^{\rm obs}_{\alpha,B}}{2},
\label{eq:g_est_massey}
\end{equation}
where $e^{\rm obs}_{\alpha,A}$ and $e^{\rm obs}_{\alpha,B}$ are the observed
ellipticities of respectively two orthogonal galaxies, whose
intrinsic ellipticities cancel each other out exactly, $e^{\rm
I}_{\alpha,A} = - e^{\rm I}_{\alpha,B}$ for both $\alpha = 1,2$.

The shear bias is then estimated from a linear fit of $g^{\rm obs}_\alpha$ as a
function of $g_\alpha$. This estimator is an improvement over the simple linear
fit reviewed in the previous section, with reduced contribution from shape
noise. However, the observed ellipticities in the absence of shear do not
cancel each other out in general, due to various effects. First, the stochasticity of the two
(assumed to be independent) ellipticity measurements means that $e^{\rm
obs}_{\alpha,A} + e^{\rm obs}_{\alpha,B}$ is a random variable with
non-zero dispersion. We model this dispersion in
Sect.~\ref{sec:err_lin_fit_orth}. Second, ellipticity bias can be different between the orthogonal pairs. Ellipticity bias can be defined from a linear fit between observed and true ellipticities (see Eq. 1 from \cite{Pujol2017}) when a true ellipticity can be defined, and it depends on the galaxies' orientation,
either with respect to the pixel coordinate system or to the PSF \citep{Pujol2017}. This can cause the
estimated shear of orthogonal pairs to be biased with respect to $g_\alpha$
\citep{Kacprzak2012,Pujol2017}. Third, selection effects can break the symmetry
if one of the two galaxies is missed. This selection can occur at the detection
level or the shape measurement stage, both of which can fail for one of the two
objects. This could be due to a dependence on the relative orientation of the
galaxy with respect to the PSF, or random noise fluctuations in particular in
the low-SNR range. Fourth, when accounting for galaxy weights, the ellipticity
cancellation is broken.

A generalization of this method consists in simulating sets of $n$ galaxies on a
ring with constant $|\bm e^{\rm I}|$, rotated uniformly such that their
mean intrinsic ellipticity is zero \citep{2007AJ....133.1763N}. The case with
$n=2$ corresponds to the case of orthogonal pairs discussed above. In
Sect.~\ref{sec:err_lin_fit_orth} we show that increasing $n$ beyond $n=2$ does
not reduce the shape-noise contribution to the shear bias
estimator.

\section{Error estimation}
\label{sec:error_estimation}

In this section we study and compare the precision of the
different shear bias estimators. In this section, a latin index of shear,
ellipticity, bias, and so on\ indicates a galaxy number from a population. The figures shown in this section are obtained from the simulated images described in
Sect.~\ref{sec:data}, and only serve for a visualization of our method. Their quantitative analysis is left for Sect.~\ref{sec:results}.

\subsection{Our method: Shear bias estimation reducing measurement noise}

Each galaxy $i$ with properties $\bm P_i$ has a shear response $\mat R_i$
estimated as described in Sect.~\ref{sec:our_method}, from different sheared
versions of the original simulated galaxy image with the same noise
realization. The response $\mat R_i$ depends deterministically
on $\bm P_i$, given by the input parameters of the simulated image, the PSF,
and stochastically on the random processes of the image realization. The
latter in our case is a simple Gaussian pixel noise realization, but we can
 include other effects such as Poisson noise and cosmic rays. The effects
on $\mat R$ from this stochasticity can be measured by repeatedly estimating
$\mat R_i$ for fixed $\bm P_i$ with different noise realizations. This provides
us with samples from the probability density function (PDF) of $\mat R_i(\bm
P_i)$. This PDF defines the uncertainty $\sigma_{{\rm N}, \alpha}$ for both
components of the estimated shear response due to stochastic effects.

\begin{figure}
\centering
\includegraphics[width=.98\linewidth]{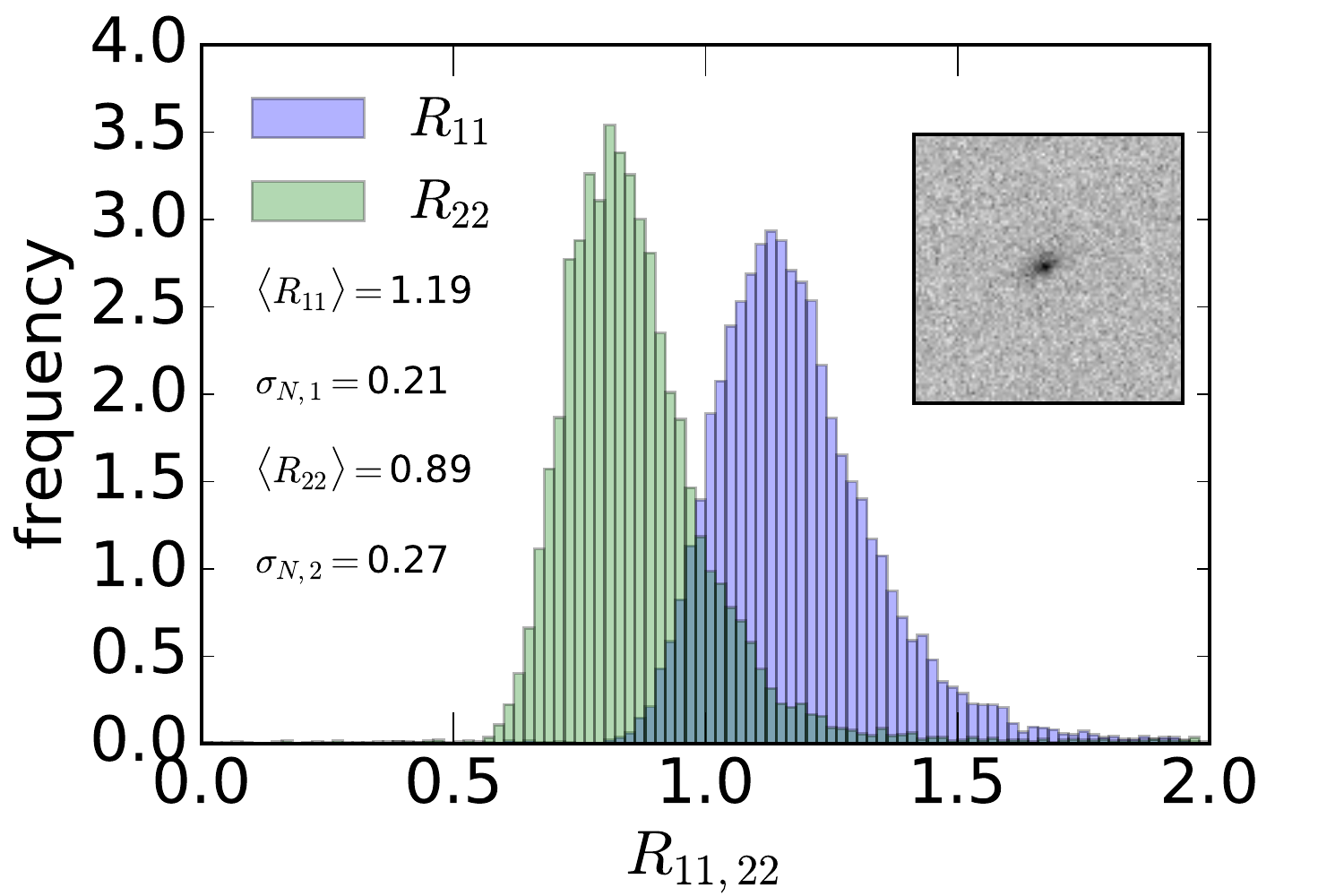}
\includegraphics[width=.98\linewidth]{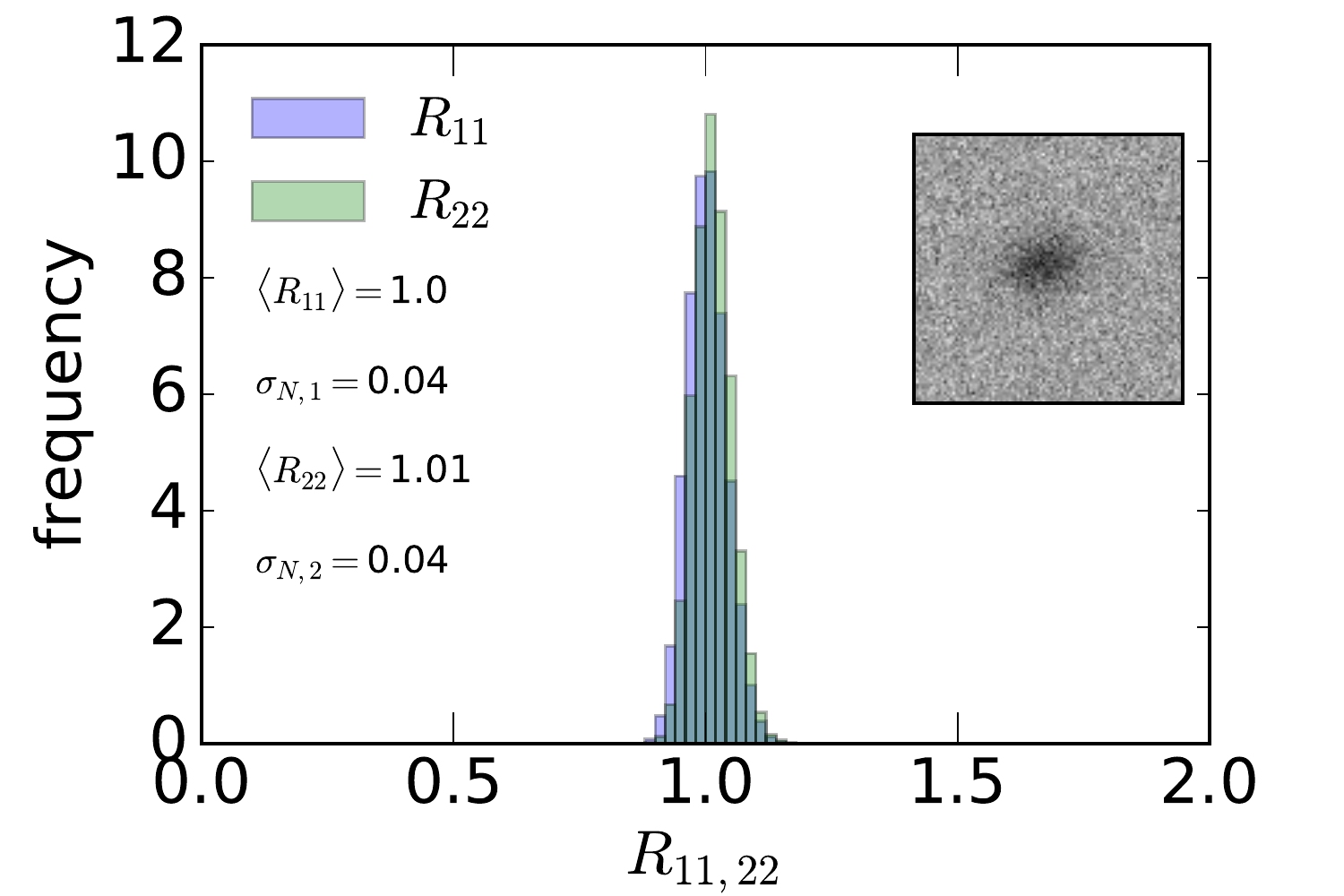}
\caption{Stochasticity of the measurement of $\mat R$ due to noise. The upper and lower panel show the distribution
  of $R_{11}$ (blue histogram) and $R_{22}$ (in green)
  for two different galaxies, respectively, shown as inlaid postage stamps, with different properties.}
\label{fig:sigma_n}
\end{figure}

In Fig.~\ref{fig:sigma_n} we show two examples of this stochasticity coming
from noise. We have measured $\mat R$ $10,000$ times for $10,000$ different
noise realizations for the two galaxies shown in the figure (see
Sect.~\ref{sec:data} for details on the simulated images and shape
measurement). As before, for each realization we do not change the noise for
the original and the four sheared versions of the image.
The mean responses $\langle \mat R_i \rangle$ depend on the galaxy properties
$\bm P_i$. In general, the response is further from $1$ for small galaxies (the top
panel) and closer for large galaxies (bottom panel), and the two response
components can be different as in the top panel. These results are consistent
with the bias results from \cite{Pujol2017}.

The dispersion for each component $\sigma_{{\rm N}, \alpha}$ of the response
depends on the noise level and on the properties $\bm P$ of the object.
The dispersion is generally larger for smaller objects. For our
shear estimation method, we only measure $R_{\alpha\alpha}$ once per galaxy,
which means that each shear response $R_{\alpha\alpha i}\bm (P_i)$ has a
stochasticity of $\sigma_{{\rm N},\alpha i}$.

Quantifying $\sigma_{\rm N, \alpha}$ allows us to estimate the number of
galaxies we need to simulate such that the stochasticity is smaller than the uncertainty we want to obtain. To meet an allowed shear bias uncertainty of $\sigma_{{\rm req},
\alpha}$, assuming that all galaxies have the same stochasticity $\sigma_{\rm
N,\alpha}$ (alternatively one can use the mean, or a worst-case value), we
would need at least $N_{\rm min} \sim \sigma_{N,\alpha}^2/ \sigma_{{\rm req},
\alpha}^2$ image simulations not to be dominated by pixel noise.
%Assuming the worst case from our simulation,
%$\sigma_{\rm N} \sim 0.27$, shown in Fig.~\ref{fig:sigma_n}, we require
%$18,000$ simulated images to beat down the stochastical (Poisson) noise.
% and $\sigma_{\rm N} \sim 0.27$ (which corresponds to the
%worst case of the small galaxy from the top panel) we would require at least
%$18000$ images to cancel the effect of noise (so, to have a complete noise
%distribution).

In the following, for the calculation of the precision of our estimator, we do not try to disentangle
the contributions from noise and galaxy properties.
Our bias estimator $m_\alpha$ for a sample of $N$ equally weighted galaxies (the application of different weights is discussed in Sect. \ref{sec:selection_weights}) is
the average of the individual shear responses,
\begin{equation}
%1 + \bm m = \langle \mat R \rangle  = \frac{\sum_{i=1}^N \mat R_i(\bm P_i)}{N} .
1 + m_\alpha = \langle R_{\alpha\alpha} \rangle  = \frac{\sum_{i=1}^N R_{\alpha\alpha i}(\bm P_i)}{N} .
\label{eq:estim_m_us}
\end{equation}
The uncertainty of the estimated response is
\begin{equation}
\sigma_{m,\alpha} = \frac{\sigma_{R,\alpha}}{\sqrt{N}},
\label{eq:Dm_meth1}
\end{equation}
where $\sigma_{R,\alpha}$ is the standard deviation of the distribution of $R_{\alpha\alpha}$.

\begin{figure}
\centering
\includegraphics[width=.98\linewidth]{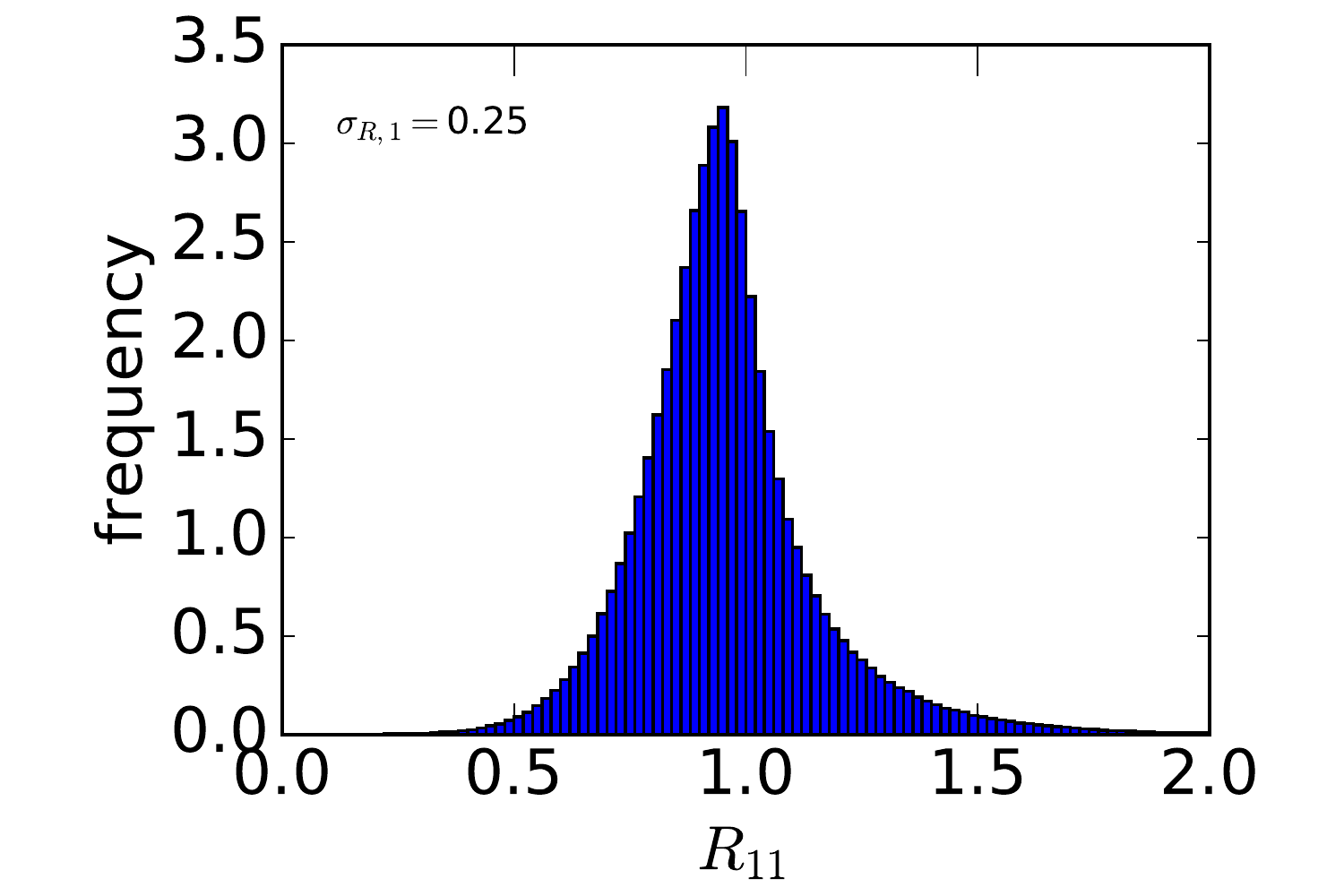}
\includegraphics[width=.98\linewidth]{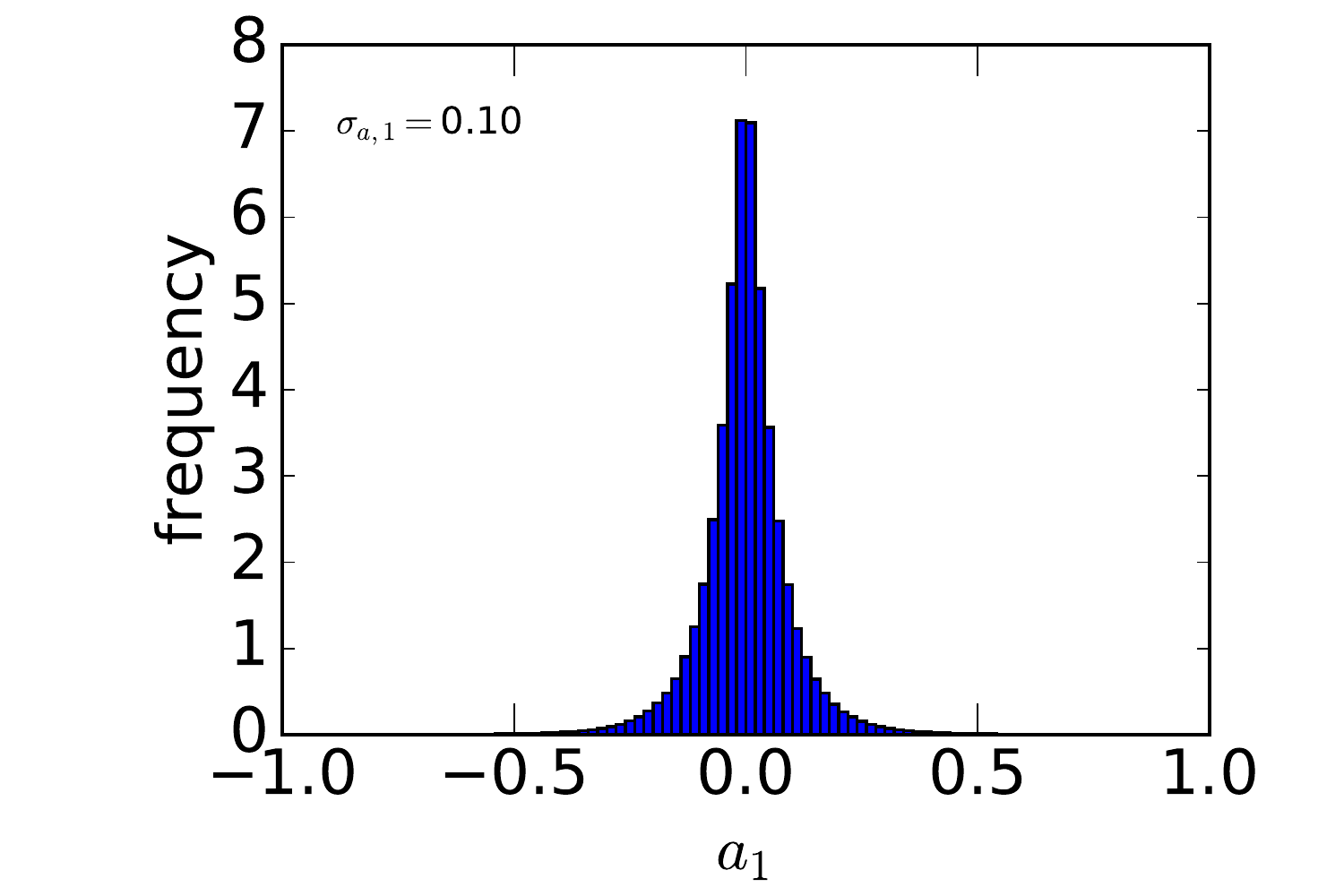}
\caption{The distribution of $R_{11}$ (top)  and $a_1$ (bottom) for the 2 million simulated galaxies. The second component of the biases shows similar distributions. }
\label{fig:sigma_b}
\end{figure}

Analogously, the additive bias is estimated as
\begin{equation}
c_\alpha = \langle a_\alpha \rangle = \frac{\sum_{i=1}^N a_{\alpha i}(\bm P_i)}{N},
\end{equation}
with uncertainty
\begin{equation}
\sigma_{c,\alpha} = \frac{\sigma_{a,\alpha}}{\sqrt{N}},
\label{eq:Dc_meth1}
\end{equation}
where now $\sigma_{a,\alpha}$ corresponds to the dispersion of the additive bias
over the galaxy population. Figure \ref{fig:sigma_b} shows the distributions of
the $R_{11}$ and $a_1$ for our sample of simulated images (see
in Sect.~\ref{sec:data}).
Only the multiplicative bias is insensitive to the ellipticity
distribution or uncertainty. The additive bias estimated using Eq. 9 is still affected by
shape noise.

\subsection{Linear fit estimation}

The observed ellipticity of a galaxy $i$ with properties $\bm P_i$ can be defined
as
\begin{equation}
e^{\rm obs}_{\alpha i} = R_{\alpha\alpha i}(\bm P_i) g_{\alpha i} +  a_{\alpha i}(\bm P_i) + S_{\alpha i}
\label{eq:m1_mult_fit2}
,\end{equation}
where $g_{\alpha i}$ is the shear and $S_{\alpha i}$ is the stochasticity
around the linear regression of the measurement for galaxy $i$ that will be
dominated by the intrinsic ellipticity $e^{\rm I}_{\alpha i}$. We write
the dependence of observed to intrinsic ellipticity as $S_{\alpha i} =
f(e^{\rm I}_{\alpha i})$ with some generic function $f$. In general, $f$
is not the identity that would represent a perfect measurement. Because ellipticity is typically larger than shear, this relation is
likely to be non-linear. When comparing the predictions with results from data,
we  only make the weak assumption that $S_\alpha$ is dominated by
$e^{\rm I}_\alpha$.

\begin{figure}
\centering
\includegraphics[width=.98\linewidth]{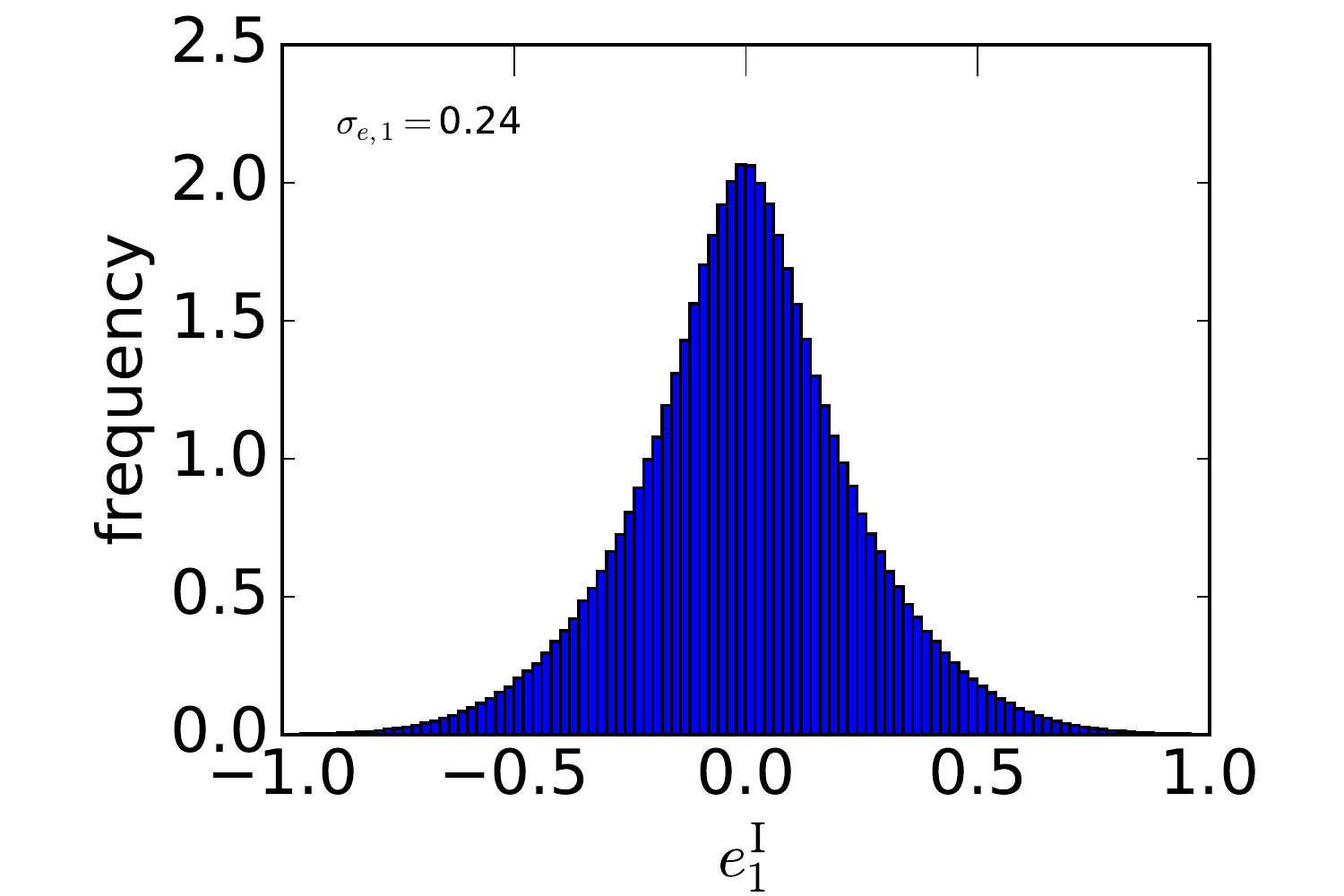}
\includegraphics[width=.98\linewidth]{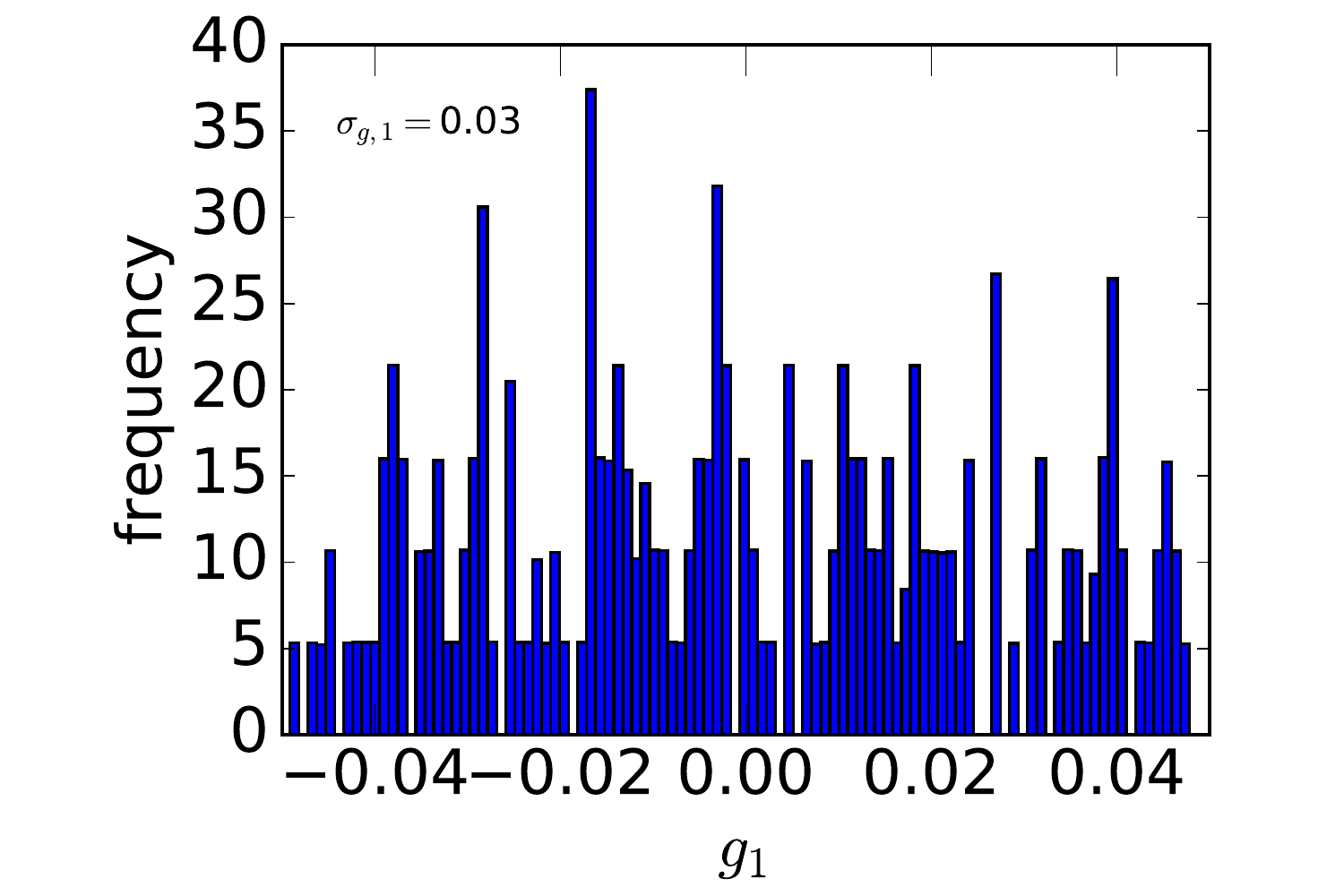}
\caption{Distributions of $e^{\rm I}_1$ and $g_1$ used for the two million simulated galaxies. The second component shows similar distributions.}
\label{fig:sigma_eg}
\end{figure}

For the linear fit to Eq. 10 we use a set of values of $g_\alpha$
and $e^{\rm I}_\alpha$, whose distributions have dispersions $\sigma_{g,\alpha}$
and $\sigma_{e,\alpha}$, respectively. In Fig.~\ref{fig:sigma_eg} we show
these distributions measured on our simulated images, which we describe in more
detail in Sect.~\ref{sec:data}.

The best values of $(1+ m_\alpha)$ and $c_\alpha$ obtained from a linear regression fit from Eq. \ref{eq:m1_mult_fit2} are given by \citep{Kenney1962} as
\begin{align}
1+ m_\alpha & = \frac{\langle (e^{\rm obs}_\alpha - \langle e^{\rm obs}_\alpha\rangle) (g_\alpha - \langle g_\alpha \rangle) \rangle}{\langle g_\alpha^2\rangle}; \\
c_\alpha & = \langle  e^{\rm obs}_\alpha \rangle -  m_\alpha \langle  g_\alpha \rangle.
 \end{align}
Assuming $\langle g_\alpha \rangle = 0$, these relations become
\begin{align}
1+ m_\alpha & = \frac{\langle ( e^{\rm obs}_\alpha -  c_\alpha)  g_\alpha\rangle}{\sigma_{g,\alpha}^2}; \\
c_\alpha & = \langle  e^{\rm obs}_\alpha \rangle.
 \end{align}
We assume that $R_{\alpha\alpha}$ and $g_\alpha$ are not correlated, which is a very good approximation since the shear bias is
linear with $g_\alpha$. Then, with
\begin{equation}
\langle (e^{\rm obs}_\alpha - c_\alpha)  g_\alpha \rangle = \langle  R_{\alpha\alpha} g_\alpha^2 +  S_\alpha  g_\alpha \rangle = \langle  R_{\alpha\alpha} \rangle  \sigma^2_{g,\alpha}  + \langle S_\alpha  g_\alpha \rangle,
\end{equation}
we find
\begin{equation}
1 + m_\alpha = \langle R_{\alpha\alpha} \rangle + \frac{\langle  S_\alpha g_\alpha \rangle}{ \sigma^2_{g,\alpha}}.
\label{eq:lin_est_m_val}
\end{equation}

The estimated $m_\alpha$ is consistent with our method if $\langle
S_\alpha g_\alpha \rangle = 0$. A correlation between these two quantities
would effectively modify the slope of the distribution of Eq.
\ref{eq:m1_mult_fit2}, resulting in a biased estimate of $m_\alpha$. For our method this condition does not need to be fulfilled.
%any of these conditions are required for our method.

We can estimate the error $\sigma_{m, \alpha}$ on $m_\alpha$ via simple
Gaussian error propagation assuming that the uncertainties in $R_{\alpha\alpha
i}$ and $S_{\alpha i}$ are uncorrelated.
This assumption would be violated if the shape estimator has a shear bias
that depends on ellipticity.
%If our shape estimator has an
%ellipticity dependence of shear bias, $R_{\alpha\alpha i}$ and $S_{\alpha i}$
%would be correlated and the following expressions would only be an
%approximation.
We test our assumptions and approximations in
Sect.~\ref{sec:results}, where we compare the numerical predictions with
measurements from simulated images. The sensitivity of the bias with respect to
these two quantities is
\begin{equation}
\left(\frac{\partial  m_\alpha}{\partial R_{\alpha\alpha i}}\right)^2 = \frac{1}{N^2}
\; ;\quad
%\label{eq:partial_m_r}
%\end{equation}
%%
%and
%%
%\begin{equation}
\left(\frac{\partial  m_\alpha}{\partial  S_{\alpha\alpha i}}\right)^2 = \frac{ g^2_{\alpha\alpha i} }{N^2 \langle  g_\alpha^2 \rangle^2} .
\label{eq:partial_m_s}
\end{equation}
Replacing for simplicity the individual galaxies' dispersions $ \sigma_{R, \alpha i}$
and $\sigma_{S, \alpha i}$ by the mean values, we get
\begin{align}
\sigma_{m, \alpha} & = \sqrt{\sum^N_{i=1} \left( \left(\frac{\partial  m_\alpha}{\partial  R_{\alpha\alpha i}}\right)^2  \sigma_{R,\alpha}^2 +  \left(\frac{\partial  m_\alpha}{\partial S_{\alpha i}}\right)^2  \sigma_{S,\alpha}^2\right)} \\
       & = \frac{1}{\sqrt{N}} \sqrt{  \sigma_{R,\alpha}^2 + \frac{ \sigma_{S,\alpha}^2}{ \sigma_{g,\alpha}^2} } .
\label{eq:Dm_m2b}
\end{align}

Compared to Eq. 7 this expressions shows the additional
term $\sigma_{S,\alpha}^2 / \sigma_{g,\alpha}^2$. In most scenarios this is
indeed the dominant term for the bias dispersion, which is the main reason why
the linear fit achieves a much lower precision in bias estimation compared to
our method.

The uncertainty on the additive bias comes directly from the dispersion in the stochasticity,
\begin{equation}
\sigma_{c,\alpha} = \frac{ \sigma_{S,\alpha}}{\sqrt{N}} .
\label{eq:Dc_m2b}
\end{equation}

\subsection{Linear fit with shape-noise suppression}\label{sec:err_lin_fit_orth}

 Here we estimate the uncertainty of the shape-noise suppression estimator
(Eq. \ref{eq:g_est_massey}), which we write in a
similar way to Eq. \ref{eq:m1_mult_fit2} as
\begin{equation}
g^{\rm obs}_{\alpha i} = R_{\alpha\alpha i}(\bm P_i) g_{\alpha i} + a_{\alpha i}(\bm P_i) + S_{\alpha i}.
\end{equation}
The difference to Eq. \ref{eq:m1_mult_fit2} is that the index $i$
now denotes a pair of orthogonal galaxies. The stochasticity
depends on the sum of the observed ellipticities of the orthogonal pair,
\begin{equation}
S_{\alpha i} = \frac{f(e^{\rm I}_{\alpha i,A}) +  f(e^{\rm I}_{\alpha i,B})}{2}.
\end{equation}
In the scenario of a perfect shape estimator, the sum vanishes exactly.
However, a shape estimator typically has a non-zero ellipticity
bias,
\begin{equation}
f(e^{\rm I}_{\alpha i,X}) = (1 +  b_{\alpha i, X}) e^{\rm I}_{\alpha i, X}
,\end{equation}
for $X=A, B$, and $g_\alpha=0$. If the ellipticity bias depends on the galaxy
orientation, or the relative orientation between galaxy and PSF or shear, the two bias
values $b_{\alpha, A}$ and $b_{\alpha, B}$ are in general not equal, and we find
\begin{equation}
S_{\alpha i} =  \frac{b_{\alpha i,A} - b_{\alpha i,B}}{2}  e_{\alpha i}^{\rm I} \ne 0.
\label{eq:S_massey}
\end{equation}

We
have measured $|b_{\alpha i,A} - b_{\alpha i,B}|$ and found it can be up to $2\%$  when one of the pairs is aligned with the shear.

The shear bias uncertainties $\sigma_{m, \alpha}$ and $\sigma_{c, \alpha}$ are
computed via Eqs. \ref{eq:Dm_m2b} and \ref{eq:Dc_m2b} derived in the
previous section, but with $\sigma_{S, \alpha}$ given by the dispersion of
Eq. \ref{eq:S_massey}. This is a clear improvement, since the pre-factor
$|b_{\alpha i,A} - b_{\alpha i,B}|$ can be expected to be smaller than unity.
In addition, if the noise realization is different for each of the objects
$A$ and $B$, this measurement is stochastic
even if $\langle b_{\alpha i, A} \rangle = \langle b_{\alpha i, B} \rangle$. This stochasticity contributes to $\sigma_{m, \alpha}$ and
$\sigma_{c, \alpha}$, which we denote with
%. with the uncertainty coming from $(\bm e^{\rm obs}_A + \bm
%e^{\rm obs}_B)$ (or equivalently $(\bm b_{A,i} - \bm b_{B,i})\bm
%e^{\rm I}_{A}$). We define this uncertainty as
$\sigma_{e_{\rm obs}, \alpha}$. In the general ring estimator case where we
simulate $n$ rotated copies of each galaxy to suppress shape noise, with
$\sum_{j=1}^n \bm e_j^{\rm I} = 0$, we can write

\begin{equation}
S_{\alpha i} = \sum_{j=1}^n e^{\rm obs}_{\alpha ij} .
\end{equation}
Keeping the total number of galaxies used in the linear fit constant, which is now $N/n$, we get
\begin{equation}
\sigma_{m, \alpha} = \frac{\sqrt{n}}{\sqrt{N}} \sqrt{ \sigma_{R, \alpha}^2
  + \frac{\sigma_{e_{\rm obs}, \alpha}^2}{n \sigma_{g, \alpha}^2} } = \frac{1}{\sqrt{N}} \sqrt{n \sigma_{R, \alpha}^2
  + \frac{\sigma_{e_{\rm obs}, \alpha}^2}{\sigma_{g, \alpha}^2} },
\label{eq:sigma_m_n}
\end{equation}
and
\begin{equation}
\sigma_{c, \alpha} = \frac{\sigma_{e_{\rm obs}, \alpha}}{\sqrt{N}}.
\end{equation}
We can see that forcing shape-noise suppression
gives a more precise $m$ than the simple linear fit as far as $
\sigma_{e_{\rm obs}, \alpha} \lesssim \sigma_{e, \alpha}$.
We also see that $\sigma_{c, \alpha}$ does not depend on the number of galaxies used
for the shape-noise suppression, but $\sigma_{m, \alpha}$ increases with $n$. However, in our derivation we neglected the
 higher-order contributions in the shear
estimator (Eq. \ref{eq:g_relation}), which decrease with $n$. In this paper we do not quantify the optimal $n$ that minimizes both contributions, since the second term in Eq. \ref{eq:sigma_m_n} dominates over these other two quantities. In conclusion, $n$ does not significantly affect $\sigma_{m, \alpha}$.

%We have measured the best-fit parameters of the following expression:
%\begin{equation}
%e^{\rm obs}_i = c_i +(1+m_i)g_i + (1+m_{e,i})e^I_i
%\end{equation}
%for $i = 1,2$, being $m_i$ the shear bias and $m_{e,i}$ the ellipticity bias. We have found $c_1 = -59\times 10^{-4}$, $m_1 = -0.049$ and $m_{e,1} = -0.124$ for our sample of galaxies (similar results are found for the second component). The fact that $m$ is different than $m_{e}$ implies that the ellipticity response is different than the shear response.
\section{Simulations}\label{sec:data}

For this analysis we used the public software package \galsim\ \citep{Rowe2015}
to generate isolated images of two million galaxies, corresponding to the
Control-Space-Constant branch of the GREAT3 challenge
\citep{Mandelbaum2014,Mandelbaum2015}. The images are organised into $200$
fields, each field with a unique PSF and shear (both constant for each field).
The galaxy light distribution follows either  a single S\'ersic profile or a de
Vaucouleurs bulge plus exponential disk.

Each galaxy is simulated twice, the second one being rotated
by $90$ degrees with respect to the first one to achieve shape-noise
suppression. For more details about the simulated images, we refer the reader
to \cite{Pujol2017} as well as \cite{Mandelbaum2014}.
This set of simulations are used for the linear fit methods, with (Sect.~\ref{sec:lin_fit_orth})
and without (Sect.~\ref{sec:linear_fit}) shape-noise suppression. For the latter, we average the observed
ellipticity of all galaxies for a given shear $g$, not specifically accounting for the orthogonal pairs when calculating the error bars (so we do not keep the galaxy pairs in the same jackknife subsamples).
This results in a mean ellipticity in each bin close to zero, but does not reduce the scatter due to
the intrinsic shape noise.

For our estimations of $\mat R$ as described in Sect.~\ref{sec:our_method}, we
simulate the two million galaxies three times, with two sheared values drawn from the
cases $\bm g = (\pm 0.02, 0)$, $\bm g = (0, \pm 0.02)$. The two values chosen
have to be different in both components in order to be able to estimate
$R_{11,22}$. Since both components of $\bm g$ change for each of the shear
versions, the estimation of $R_{\alpha \alpha}$ is affected by the non-diagonal
terms as follows:
\begin{equation}
\Delta e^{\rm obs}_\alpha = R_{\alpha 1}\Delta g_1 + R_{\alpha 2}\Delta g_2.
\end{equation}

Our estimation of $\langle R_{\alpha \alpha}\rangle$ is unbiased as far as
$\langle \Delta g_\beta \rangle = 0$, for $\beta \ne \alpha$, over the entire
sample. This is the case since we choose the sign of the shear changes at
random. We can also measure the non-diagonal terms of $\mat R$ by using three
images with shear values from $\bm g = (0, \pm 0.02)$, $\bm g = (\pm 0.02, 0),$
and $\bm g = (0, 0)$ (see Appendix \ref{sec:robustness_m1} for more details).

Galaxy shapes are obtained with the
method from \cite{Kaiser1995} (\ksb), using the publicly available code \shapelens\ \citep{Viola2011}. This
method estimates the ellipticity of the objects from the
surface brightness moments
\begin{equation}
Q_{\alpha\beta} = \frac{\int{{\rm d} ^2 x I({\vec x}) W({\vec x}) x_\alpha x_\beta} }{\int { {\rm d}^2 x I({\vec x}) W({\vec x}) }},
\end{equation}
defining the ellipticity as
\begin{equation}
\bm e = e_1 + {\rm i} e_2 = \frac{Q_{11} - Q_{22} - 2 {\rm i} Q_{12}}{Q_{11} + Q_{22} }.
\end{equation}
The implementation details of the shape measurement algorithm are not very relevant for this
paper, and we refer the reader to \cite{Pujol2017} where we used the same methodology.

\section{Results}\label{sec:results}

\begin{figure}
\centering
\includegraphics[width=.98\linewidth]{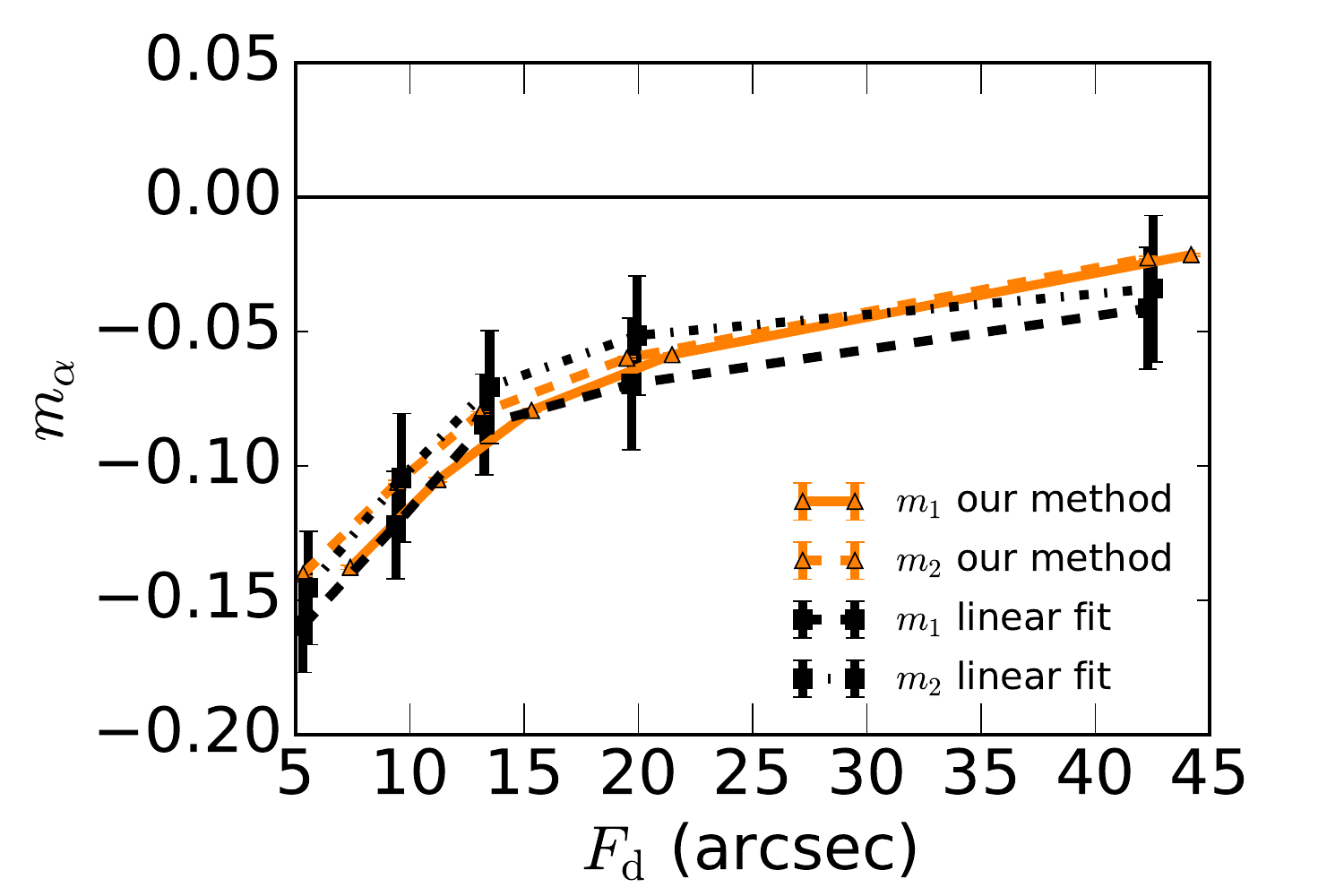}
\includegraphics[width=.98\linewidth]{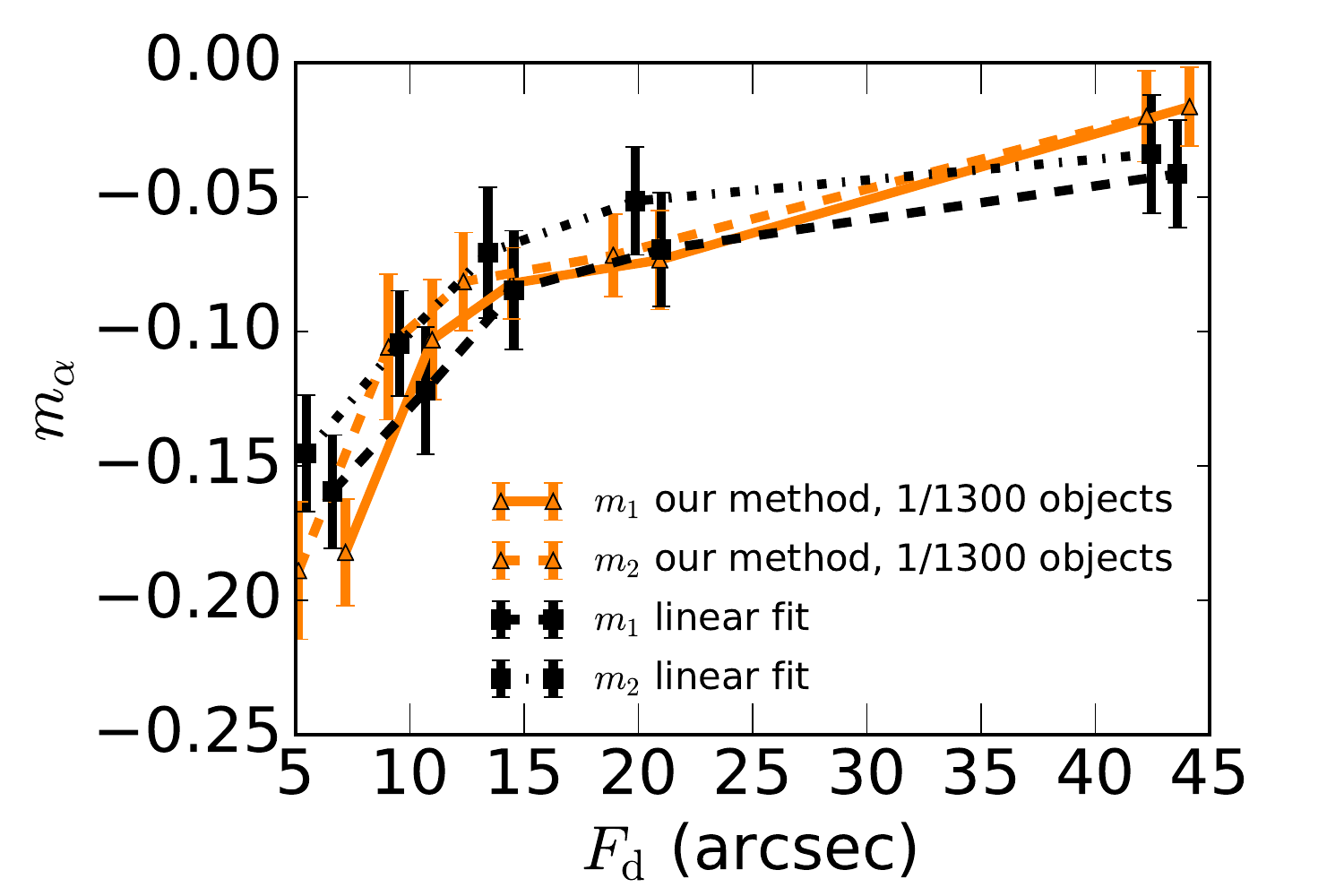}
\caption{Multiplicative shear bias as a function of the disk flux $F_{\rm d}$, measured with our method
(black lines) and (in orange) from the linear fit to Eq. \ref{eq:g_relation}. Solid (dashed) lines correspond to $m_1$ ($m_2$). The top panel shows the results using the same number of object for both methods. In the bottom panel, only 1/1300 objects have been used for our method.}
\label{fig:b_vs_snr_comp}
\end{figure}

\begin{figure}
\centering
\includegraphics[width=.98\linewidth]{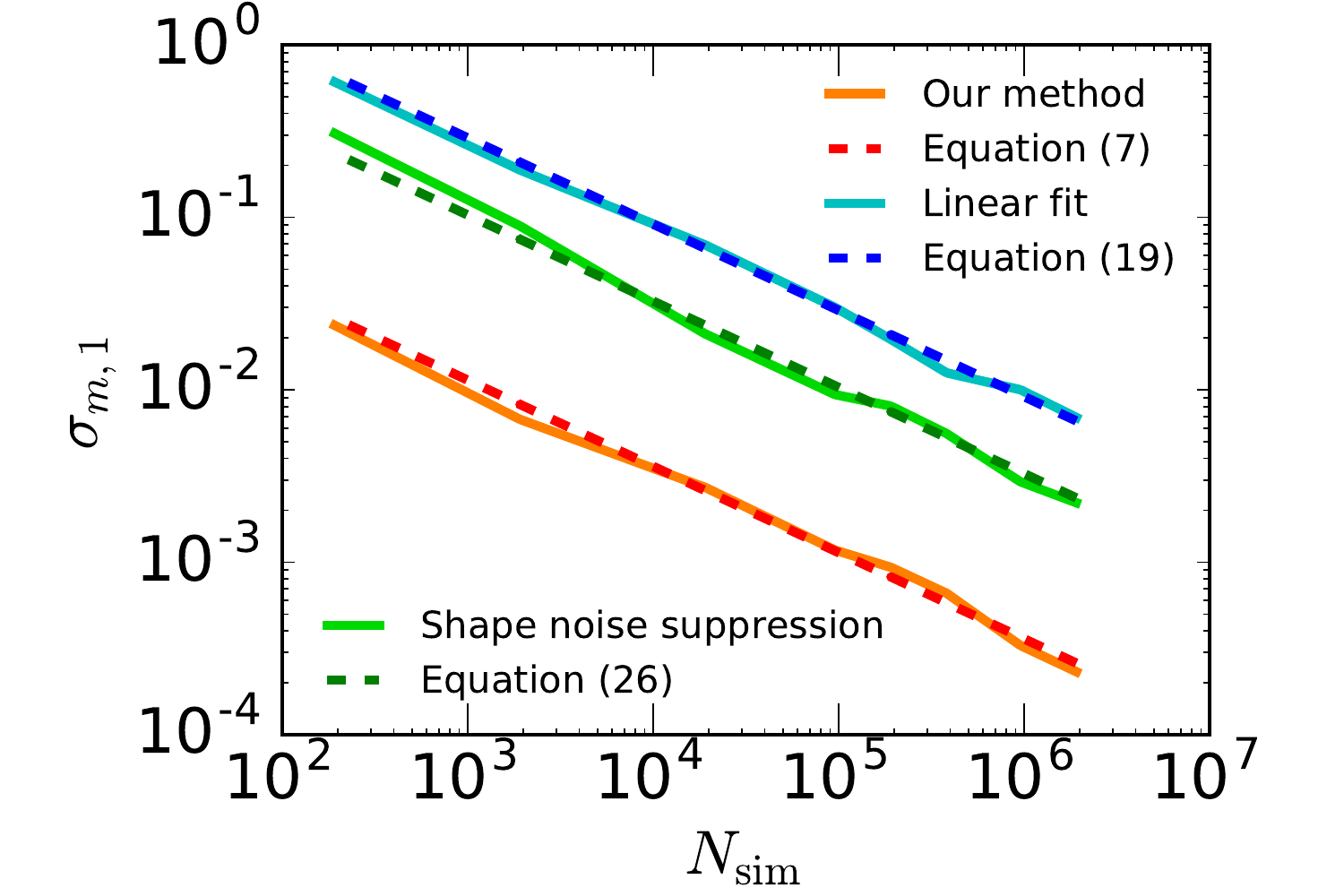}
\includegraphics[width=.98\linewidth]{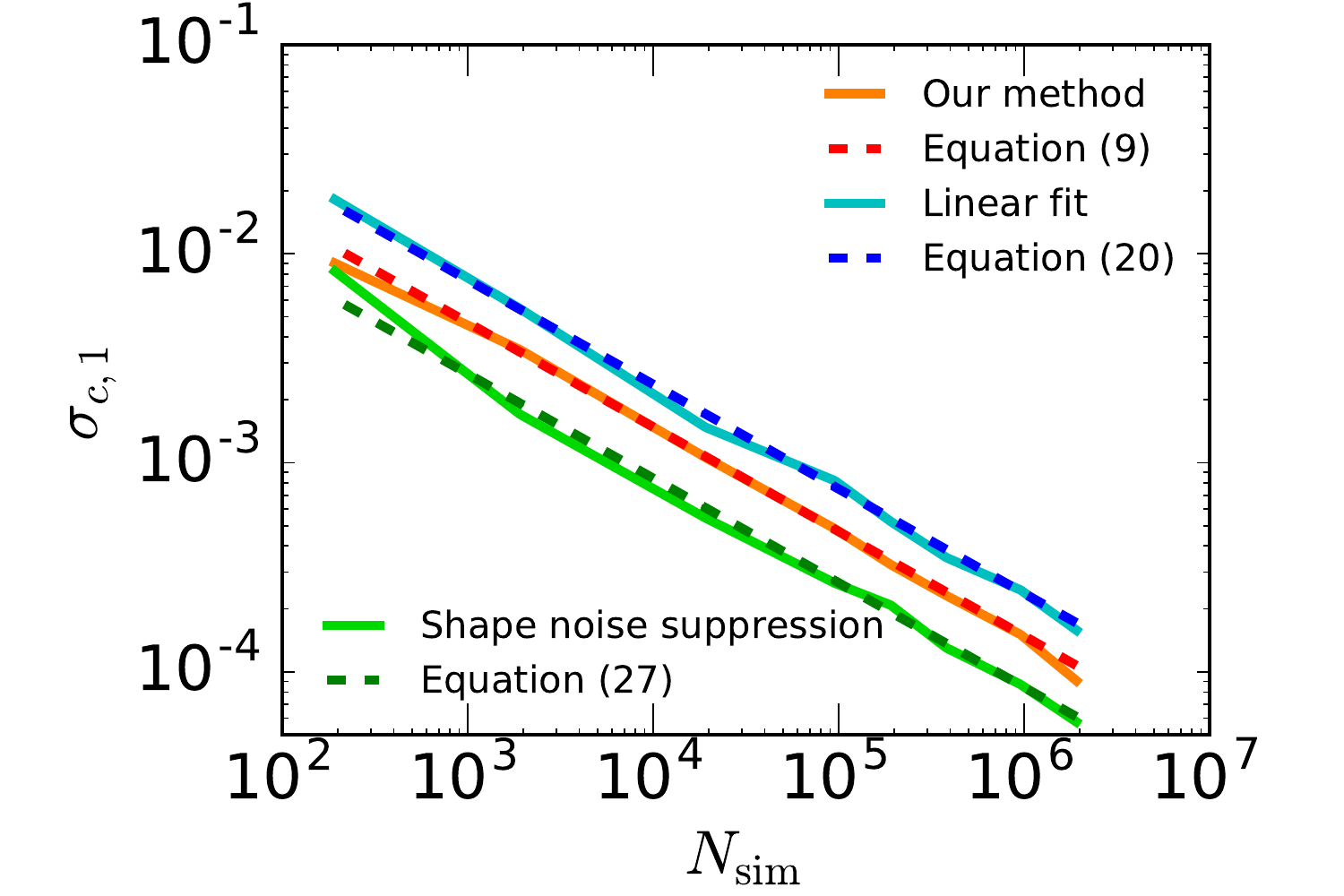}
\caption{RMS of the multiplicative (top panel) and additive (bottom panel) shear bias.
We compare our method (red/orange lines) to the linear fit with (green) and without (cyan/blue) shape-noise
suppression. The solid lines are measurements from the numerical simulations.
Dashed lines show the analytical predictions derived in Sect.~\ref{sec:error_estimation}.
}
\label{fig:Dm_meth1}
\end{figure}

In the top panel of Fig.~\ref{fig:b_vs_snr_comp} we compare the shear bias obtained with our
method to the linear fit technique. As an example of galaxy property we use the
input disk flux $F_{\rm d}$ of the simulated bulge+disk galaxies. We show that both methods give consistent results when using
all two million galaxies. However, our method estimates the biases with a
significantly better precision. The location of the points on the $x$-axis
corresponds to the centre of the $F_{\rm d}$ bins. In addition to a small shift
that we apply for an easier visual comparison, the bin centres for our method
in the lower panel are modified, since the galaxies are  now a random subsample.
It is remarkable that when using all two million galaxies, the curves of
$m_1$ and $m_2$ for our method are almost identical.

We quantify the precision of the different shear bias estimation methods in
Fig.~\ref{fig:Dm_meth1}. as a function of the number of simulated galaxies $N_{\rm sim}$. We create
different random subsets of galaxies with size $N_{\rm sim}$, and measure for each
subset the shear bias for the three methods as described in
Sect.~\ref{sec:shear_bias_measurement}. We compute the RMS for each subset by
jackknife resampling of the input galaxies for all methods, using $50$
subsamples (other numbers of subsamples have given the same results).

% where we compare the
%accuracies $\bm \sigma_m$ (top panel) and $\bm \sigma_c$ (bottom panel)
We compare these uncertainties as measured from the simulations to the
numerical predictions derived in Sect.~\ref{sec:error_estimation}. For the
latter, we measure the parameters $\sigma_{R, \alpha}$, $\sigma_{a, \alpha}$, $
\sigma_{S, \alpha}$,  $\sigma_{e^{\rm obs}, \alpha}$, and $\sigma_{g, \alpha}$
directly from the simulations, as illustrated in
Figs.~\ref{fig:sigma_b} and \ref{fig:sigma_eg}.
%obtained from the data for the different methods with the theory predictions.
The amplitude and $N_{\rm sim}^{-1/2}$-dependence of the uncertainty measured from the
data shows excellent agreement with the analytical calculations for all three
methods. This suggests that the assumptions we made to derive these
expressions are valid for the system and regime studied here.
For the linear
fit predictions, we set $\sigma_{S, \alpha} = \sigma_{e, \alpha}^{\rm I}$, assuming
that stochasticity $S_\alpha$ is entirely determined by the intrinsic ellipticity.
For
the linear fit with shape-noise suppression, we measure $\sigma_{e^{\rm out}, \alpha}$ directly from the distribution of the sum of observed ellipticities
of the orthogonal pairs, $(e^{\rm obs}_{A, \alpha} + e^{\rm obs}_{B, \alpha})/2$.

%
%The solid red lines show the data results obtained from our method, showing a
%good consistency with the theory predictions from equations (\ref{eq:Dm_meth1})
%and (\ref{eq:Dc_meth1}), represented by dashed black lines. The solid cyan
%lines show the data results obtained from the linear fitting method, and they
%are compared with their theory predictions from equations (\ref{eq:Dm_m2b}) and
%(\ref{eq:Dc_m2b}) (dashed blue lines). Finally, the data results from the
%linear fit with a  shape noise cancellation using orthogonal pairs is shown in
%\WORK{make lines} and compared to the theory predictions from equations
%\WORK{make equations} (dashed green lines).
%All the methods show an excellent
%agreement between the theoretical predictions and our numerical results from
%the image simulations, suggesting that the assumptions we have made to derive
%these expressions are valid for the system and regime studied here. For the
%model prediction of the linear fit (dashed blue lines) we assumed that $\bm
%\sigma_S = \bm \sigma_e^{\rm I}$, suggesting that $\bm S$ is entirely dominated
%by $\bm e^{\rm I}$ in our case.

Our method has a much higher precision on the multiplicative shear bias
estimation. Compared to the linear fit, $\sigma_{m, \alpha}$ for our method is
smaller by a factor of $35.9$. This means that for this study our method
requires $35.9^2/n^\prime \sim 1300/n^\prime$ times fewer simulated images to
obtain the same precision, where $n^\prime$ is the number of sheared versions
used for each object.
In Fig.~\ref{fig:Dm_meth1} we show our method with $n'=2$, where we used shear values symmetrically distributed around zero, but similar results have been found for $n'=4$.

We demonstrate the high precision of our method in the bottom panel of
Fig.~\ref{fig:b_vs_snr_comp}, where we estimate the shear bias as in the top
panel, but now for our method with only a fraction of $1/1300$ of the objects,
chosen at random. The results are consistent in both mean and error bars,
demonstrating that our method reaches the precision of existing
methods with three orders of magnitude fewer simulations.
We note that some of the noise in the data points for our method comes from the
more sparsely sampled galaxy properties in each bin.
In the case of Euclid, with a global requirement of $\sigma_{m, \alpha} < 2
\times 10^{-3}$ one needs at least $2\times 10^7$ images for the linear fit
method, but only $\sim 10^4$ for our method according to this study.
%Note that here the values of our methods show
%slightly noisier results. This is because for our method only $\sim 10,000$
%objects have been used, so the distribution of properties of the objects is
%more discrete and hence noisier and not identical to the $2,000,000$ objects.

The ratio of the RMS between the two methods is approximately
$\sigma_{e, \alpha} /(\sigma_{R, \alpha} \sigma_{g, \alpha}) > 1$. The
quantities $\sigma_{e, \alpha}$ and $\sigma_{g, \alpha}$ in the
simulation need to be chosen to match expectations from cosmology and galaxy
morphology. Given some basic survey characteristics such as redshift and
wavelength coverage, and the survey selection function, these fundamental
quantities are fixed. The dispersion $\sigma_{R, \alpha}$ , however, strongly
depends on instrumental effects such as the PSF size and on the shape
estimator. In this study we used a \ksb\ method to measure the shapes on
GREAT3-CSC-like images, and we expect $\sigma_{R, \alpha}$ to change when using
other simulations and shape estimators. This provides a strong motivation
to choose or develop a shape measurement method that minimizes this dispersion,
and therefore minimizes the number of required simulations for calibration.

Applying the shape-noise suppression with orthogonal pairs improves the
precision with respect to the simple linear fit by a factor of $\sim 2.8$ for
the measurements of both the multiplicative and additive bias. This improvement
reduces by a factor of $\sim 8$ the number of simulated images required for the
same level of precision. We note, however, that each galaxy needs to be simulated twice for the shape-noise suppression.  This is consistent with the factor of $\sim 9$ found in \cite{Conti2016}, where they used $n=4$ for the shape-noise suppression.

Comparing our method to the linear fit with shape-noise suppression,
we obtain an
improvement of a factor of $12.8$ for the multiplicative shear bias. This implies
that for the same level of precision, we can reduce the number of simulated
images required by a factor of $12.8^2/n' \sim 164/n'$.
% with respect to the
%method with orthogonal pair shape noise suppression.
%, where again $n'$ is the
%number of sheared versions used per image for our method.

When comparing the additive bias precision, our method shows a factor of $2.26$
improvement with respect to the linear fit, and a factor $0.56$ with respect to
the shape-noise suppression. Shape-noise suppression performs better
because the additive bias is the average ellipticity over all simulated images, while for
our method only $1/n'$ images are used. In principle we could estimate $c_\alpha$
with $n'=1$ (e.g.~using only the original image). This would, however, unfairly not count
the $n^\prime >1$ images we have to simulate to measure $m_\alpha$.
We conclude that a similar precision is obtained for both our method
and shape-noise suppression when estimating additive shear bias.

%For the method with shape noise suppression we have assumed $\langle b_{\alpha i, A} \rangle = \langle b_{\alpha i, B} \rangle$, but we can have different ellipticity
%bias for the orthogonal pairs, since their alignments with respect to the shear or PSF
%is different in general, and thus their observed ellipticity modulus will be different
%both ob
%of the objects with
%respect to the shear will be different and hence so will be their shapes.
%We
%have measured the differences in the ellipticity bias as a function of the
%relative orientation with the shear, finding that orthogonal pairs can have
%differences in their ellipticity biases up to $2$ percent. This would also
%reduce the precision of this method.
%Our numerical results suggest that this effect is negligible, although its impact depends on the image
%specifications and the shape estimator used.

%Assuming the ellipticity biases to be the same as the shear bias would also be
%incorrect. For our image simulations we have found an average shear bias of $m_1
%= -0.049$ and an average ellipticity bias of $b_1 = -0.124$ (similar results are found for the 2nd component). This large difference between those biases confirms the findings from \cite{Pujol2017}.

%We also remind the reader
%that any selection effects or weights affecting differently the
%orthogonal pairs (which can be expected since their orientation relative to the
%shear is different and then so their global shapes) would increase the
%uncertainty of the method significantly.

\section{Discussion and applications}
\label{sec:applications}

The method presented here is a clear improvement on the precision of the shear
bias estimation in simulations with respect to the standard linear fit of
Eq. \ref{eq:g_relation}. It is also more precise compared to the
linear fit with shape-noise suppression via pairs of orthogonally aligned galaxies.
In the following we
discuss potentially useful applications to improve shear bias analyses with
simulated images.

\subsection{Shear bias validation and calibration}

One of the interests of measuring shear bias in simulations is to validate or
calibrate the performance of a shear estimation algorithm. In the case upcoming
surveys such as Euclid, LSST, or WFIRST the requirements concerning the knowledge of the
additive and multiplicative bias imply the generation of a very large volume of
simulations, which is computationally very challenging. Our method  allows the
saving of significant computational efforts to reach these requirements. In our case study, we require $2$ to $3$  orders of magnitude fewer images to reach the same
precision as common approaches, although the exact factor depends on the shear
estimator algorithms and the image and survey specifications.

\subsection{Selection biases and weights}\label{sec:selection_weights}

Shear bias from selection effects has been found to be of the same order of
magnitude as those induced by the shape measurement process \citep{Conti2016,2017arXiv171000885M}.
Such biases arise when the galaxy selection function depends on the shear. This
is for example the case when detection or shape measurement fails for galaxies
that are very elliptical, or aligned with the PSF. Such selection effects also arise
by imposed, necessary cuts on galaxy properties such as the signal to noise ratio (SNR) or size, which can favour
certain shear values.
The resulting shape catalogue
then samples the underlying shear field in a non-representative way, which induces biases on
the estimated shear if uncorrected.

Our method does not require shape-noise suppression via tuples of
galaxies, and is therefore particularly useful when selection effects and weights are to
be simulated and studied. Weights can be applied to the simulated galaxies following an arbitrary
distribution, to study the impact on shear bias. When weights are given to the galaxies, the shear bias estimators become

\begin{equation}
  m_\alpha = \frac{\sum_{i = 0}^N w_i R_{\alpha\alpha}}{\sum_{i = 0}^N w_i}
\end{equation}
\begin{equation}
  c_\alpha = \frac{\sum_{i = 0}^N w_i a_{\alpha}}{\sum_{i = 0}^N w_i},
\end{equation}
where $w_i$ represents the weight of the $i$th galaxy and $N$ is the total number of galaxies.

Selection effects that are correlated with the shear can be studied as
proposed by \cite{Sheldon2017}: we calculate the mean response from
Eq. \ref{eq:ind_bias_estim}  by first averaging the ellipticities of both sheared
samples before taking the difference and dividing by the small shear,
\begin{equation}
  \left\langle R_{\alpha\beta} \right\rangle \approx
  \frac{\left\langle e^{\rm obs, +}_\alpha \right\rangle - \left\langle e^{\rm obs, -}_\alpha \right\rangle}{2 \Delta g_\beta} .
  \label{eq:ind_bias_estim_mean}
\end{equation}
Now, the two galaxy samples giving rise to the mean observed ellipticities
$\left\langle e^{\rm obs, +}_\alpha \right\rangle$ and $\left\langle
e^{\rm obs, -}_\alpha \right\rangle$, respectively, are not only
different because of their shear. In addition, a given selection criterium (e.g.~a
minimum SNR) is applied to the two sheared samples. If the applied shears
modify the selection, this results in different mean sample ellipticities, and
the selection-induced shear bias translates into the shear response
(given by Eq. \ref{eq:ind_bias_estim_mean}).

This shear bias estimator, however, does not account for selection effects
that affected the shear response estimation of the originally selected galaxies. It can happen that the shear response estimation fails because, although the original galaxy is well detected and measured, this is not the case for the sheared version of the image. These selection effects are undesired, since they create
additional, spurious selection biases. Such differential detections or
shape measurement successes or failures is rare however,
since the shear is very small and thus the images are very similar.
Such occurrences can be further reduced: if an image sheared by a value $\bm g$
cannot be measured, for example~because its increased observed ellipticity pushed it under the SNR threshold, the opposite shear
$-\bm g$ (e.g.~making the galaxy rounder) should not affect the measurement
success. In the case of $n^\prime = 2$ images per original galaxy, we are free
to choose the sign of the shear as long as the average shear is zero, largely avoiding
such selection-induced measurement failures. If the problem only comes from the detection process, another solution can be applying the detection process to the original images and assume the same detection for the sheared versions of the images.

\subsection{Shot noise}

In the methodology described the shear response is estimated from sheared versions of images keeping the noise realizations fixed. This can be generalized to different random or stochastic effects such as cosmic rays or Gaussian noise. However, shot noise is a random process that depends on the flux of the image. Because of this, sheared versions of the same galaxy cannot have  exactly the same shot noise realization. Our case is not affected by this, since noise is purely Gaussian, and we can expect other cases to be also insensitive to shot noise, but this is not always the case.

Exploring alternatives to treat shot noise with our method is beyond the scope of this paper, but we propose several options. First, in some cases approximating shot noise with a Gaussian noise can be enough for the required precision of the analysis, which again can be treated as described in this paper. Second, we can keep the random shot noise realization of the original image and rescale it with the changes in the flux produced in the sheared versions. A study should be done to test possible systematics coming from this approach. Finally, we can change the shot noise realization for each of the sheared image versions, but keep the other random processes fixed. This will degrade the precision of the method depending on the contribution of shot noise with respect to the other processes, but it should converge to the same results. In the worst case scenario where the shear response depends completely on shot noise, the precision of the method would be the same as for the method with orthogonal pair shape-noise suppression.

\subsection{Individual shear responses}

Studying the shear response as a function of galaxy properties for individual galaxies
without the need to bin or average can have advantages.
For calibration, the shear bias as a function of galaxy properties is typically modelled
as a smooth function, either parametric, for example~by fitting an analytical, multi-variate function,
or non-parametric, such as~by interpolating the (smoothed) measured bias values.

 For linear fit methods to estimate shear bias we need to compute such a function from data binned into galaxy properties. Then the average shear biases are measured for each bin. However, these average values depend on the galaxy population inside the bin, whose shear responses might not only depend on the binned properties, but also on the properties that have not been used in the binning. As a consequence, our measured shear bias dependencies are sensitive to the property distribution of the galaxy population used.
Individual shear responses and biases of simulated galaxies can further serve to learn shear calibration as a complex non-linear function of galaxy and image properties (e.g. using machine learning techniques), where no binning is needed and we can use a larger set of properties so that the function can be less dependent on the population used.

\subsection{Variable shear and response on shear statistics}
\label{sec:variable}

%So far we have simulated galaxies either with a random or with no shear, before
%applying a small constant shear.
Switching from constant to variable shear is possible by imposing a shear field on our simulation. This is potentially interesting to study the scale dependences of shear bias that could come from spatially varying effects such as the PSF variation. Similarly to the constant shear case, the shear bias is derived by computing the shear response to a small shear power spectrum perturbation. For example, with the shear drawn from a Gaussian random field with a certain power spectrum $C_\ell$, the small shear values applied to each galaxy with an intrinsic shear $\bm g$ can be arbitrary, for example~they can be proportional to $\bm g$ (see discussion in Sect.~\ref{sec:blended}).
Going a step further, the same methodology can be applied to derive the influence of shear bias on any shear statistics: the shear two-point correlation function, the shear power spectrum, peak counts,
mass maps, higher-order statistics, and so on. As an example we hereafter illustrate this with the shear two-point correlation function $\xi_\pm$.

First, as described above, we apply a shear field to the simulated galaxies, where each
Fourier-space shear coefficient $\bm \hat \gamma_{\bm \ell}$ is drawn from a normal distribution
with zero mean and variance $C_\ell$. Next, we perturb the shear field by drawing new coefficients
$\hm \hat \gamma_\ell \sim {\cal N}(0, C_\ell + \delta C_\ell)$, where we change the power spectrum
by a small amount, $\delta C_\ell$.
From the original and perturbed shear field, we compute the statistics of our choice, for example~the correlation functions
$\xi_\pm$ and $\xi_\pm^+$, respectively. The difference between both divided by the perturbation is then
the response due to the multiplicative shear bias on the correlation function, which would give us information about the spatially varying shear bias.

\subsection{Non-isolated images}
\label{sec:blended}

This analysis has been done using isolated galaxy images. To create
more realistic simulations with blended galaxy images leads to the problem that
the shape of many galaxies is measured in the presence of one or more nearby
galaxies at different redshift and therefore different shear, if the simulation
presents a realistic cosmic shear field as described in the previous section.
The same issue arises if shapes of blended galaxies are estimated jointly. The
presence of nearby isophotes of other objects is known to affect the shear
bias \citep{Hoekstra2015,Hoekstra2017}.

A common procedure to study these effects is by simulating many combinations of
blended objects and close neighbours, and measuring the impact on the shear bias
statistics over different populations. We claim that we can more
efficiently account for these effects, since our shape-noise insensitive method
does not require us to sample the large space of the distribution of $N$
ellipticities and shears $p(e_1, \ldots, e_N, g_1, \ldots g_N)$.
One of the questions to address in this situation is how to produce the
different sheared versions of the same images. Here we discuss two
possibilities:
\begin{itemize}
\item We change the shear of only one of the galaxies (the target) from the $N$-tuple of blended images. The
inconvenience is that we need to generate $N$ times more images compared to
isolated galaxies.
\item Alternatively, we can shear every member $i$ of the $N$-tuple.
This shear could be a small additive shear, $\Delta \bm g = \mbox{const}$, as
applied to isolated images in this paper. Or it could be a function of $\bm g$, such as a multiplicative factor, $\Delta \bm g_i = C \bm g_i$ with $C \ll 1 = \mbox{const}$.
In this case we would preserve the proportions between shears for galaxies at different redshifts. This function of $\bm g$ can be chosen taking into account the statistics or cosmological analysis that we want to do, as discussed in Sect. \ref{sec:variable}.

%\mk{Don't understand the following sentence:}
%In any case, the main contribution of the shear responses would come
%from the intrinsic shapes of the objects, for which intrinsic alignments can
%have a stronger impact.
%
\end{itemize}

We leave a comparison of these two approaches, and estimation of shear bias for blended objects in general, to future work,
which is beyond the scope of this paper.

\subsection{Future simulation challenges}

Adopting future versions of simulation challenges such as the GRavitational lEnsing Accuracy Testing (GREAT) series
\citep{Bridle2009,Kitching2010,Mandelbaum2014} into our method of shear bias
estimation can result in a significant decrease of required image simulations.
For GREAT3 the total data volume that had to be downloaded by the participants
was 6.5 terabyte ($10,000$ simulated galaxies $\times$ 200 fields $\times$ 20
branches). Reducing this number could result in a more accessible and faster to
process challenge.

To use our method, for each galaxy two additional sheared versions of the
same galaxy would need to be simulated with the same noise realization. The
challenge organisers would estimate the shear response for each original galaxy
via Eq. \ref{eq:ind_bias}, and apply some metric on the distribution, such as~the
mean, to evaluate the submissions.

To guarantee the blind aspect of the challenge,  all codes have to be run
on the organisers' server without direct access to the simulation by the
participants. (For testing, smaller training sets of simulations could be
provided to the teams for download.). Alternatively, the shear values applied to each
galaxy have to be random and kept hidden from the users. We note that it would be trivial for
participants to identify the sheared versions of each original galaxy since the
noise is the same for the sheared versions, even if the image order was
randomized.
To take the example of GREAT3, a similar challenge using our bias estimator could
reduce the $10,000 \times 200$ simulated galaxies for one branch to a few thousand.
If a variable shear field is to be used in the challenge, with a metric
operating on the shear correlation function or power spectrum, a similar method
as described in Sect.~\ref{sec:variable} can be employed to measure the response to a small
and variable shear.
\section{Summary}\label{sec:conclusions}

In this paper we present a new method to estimate shear bias from image
simulations. Our estimator of the multiplicative shear bias is not affected by shape noise and reduces the noise contribution from the measured shape,
removing the dominant uncertainty in bias estimation. Previous methods constrain the multiplicative and additive bias from
a linear fit of the observed average ellipticity as a function of shear.
The uncertainty of this parameter estimation is dominated by the intrinsic
ellipticity distribution.  Shape-noise suppression techniques using matched sets of galaxies with net zero intrinsic ellipticity
 improve the precision of the measurements, but are affected by
selection effects, weights, and ellipticity bias that can break the shape-noise suppression.

Our method consists in measuring the shear response and additive bias of
individual galaxy images. To that end, we simulate different sheared versions of
the same galaxy, and measure the
shear response of the image from the numerical derivative of the measured ellipticity
with respect to the shear. We also measure the additive bias for the
individual images. For each galaxy the sheared version has the same noise realization, allowing us to determine
the individual responses at a very high precision. Then, the multiplicative
and additive bias of a sample of galaxy images is obtained from the average
shear response and additive bias, respectively. This method improves the
precision of the estimation of shear bias significantly because it is not
affected by shape noise or by the stochastic uncertainty of the measured
ellipticity.

Using numerical simulations as well as analytical predictions, we quantified the uncertainty
 of the shear bias estimation for
our method as well as for linear fits.
For the multiplicative shear bias, our method provides a significant decrease in
the shear bias error of a factor of $\sim 36$ compared to the linear fit, and
a factor of $\sim 12$ if the latter is used in combination with shape-noise
suppression. The additive bias uncertainty improves by about $2.3$ over the linear fit,
and under-performs only compared to shape-noise suppression, by a factor $\sim 0.5$.

This implies that we can reduce the number of simulated images by a factor of
$\sim 1,300$  and $\sim 150$, respectively, to measure the shear multiplicative bias
with the same precision. Our method has the further advantage that it does not
need to impose shape-noise suppression, and hence it can easily be applied for analyses
where selection biases or weights play an important role.

Our method has many applications as discussed in the previous section.
In particular for shear bias calibration, we require much fewer simulated images to reach a required uncertainty,
allowing us to study more extensively the bias dependence as function of galaxy property, PSF characteristics, or noise.
It also relieves us of the potentially very severe restrictions on computing time for both simulation and
shape measurement, allowing us to simulate galaxies with higher complexity, and using
computationally expensive shape measurement techniques.
Further, it permits us to study the shear bias as a function of galaxy properties that usually have to be averaged over,
for example galaxy orientation.
We have also outlined ways forward for more complex simulation scenarios, such as variable shear, blended galaxy images,
and selection biases, which in principle pose no obstacles for our method.

\section*{Acknowledgements}
The authors would like to thank Rachel Mandelbaum, Richard Massey, Arun Kannawadi, Lance Miller, and Henk Hoekstra for very helpful comments and suggestions. We also thank the anonymous referee for the useful feedback that helped improve the paper.
AP, FS, and JB acknowledge support from a European Research Council Starting Grant (LENA-678282). AP and MK are supported by the
French national programme for cosmology and galaxies (PNCG).

\bibliography{biblist}

\appendix

\section{Robustness of Method 1}\label{sec:robustness_m1}

In this section we investigate whether a least-squares fit to more than two shear values is more accurate than the use of Eq. \ref{eq:ind_bias_estim}.
To measure $\mat R$ for each individual galaxy, we have generated copies of the same images with the different shears specified in Sect. \ref{sec:our_method}. In reality, galaxies can have other values of shear, where both shear components can be different from $0$ at the same time. Fixing the other component to $0$ can be a simplification of the estimation of $\mat R$. For this reason, here we measure the impact of different estimations of $\mat R$ using different shear values. In particular, we compare the following estimators:

\begin{itemize}
\item $R_{11}^A$: we obtain $R_{11}$ from the fit using the shear values $\bm g = (\pm 0.02, 0)$.
\item $R_{11}^B$: we obtain $R_{11}$ from the fit using the shear values $\bm g = (\pm 0.02, 0)$ and $\bm g = [0,0]$.
\item $R_{11}^C$: we obtain $R_{11}$ from the fit using the shear values $\bm g = (\pm 0.02, 0)$ and a random value of $\bm g$, with both components random.
\item $R_{11}^D$: we obtain $R_{11}$ from the fit using the shear values $\bm g = (\pm 0.02, 0)$,  $\bm g = [0,0]$ and the random $\bm g$.
\item $R_{11}^E$: we obtain $R_{11}$ from the fit using all the previous values and also $\bm g = (0, \pm 0.02)$.
\end{itemize}

\begin{figure}
\centering
\includegraphics[width=.98\linewidth]{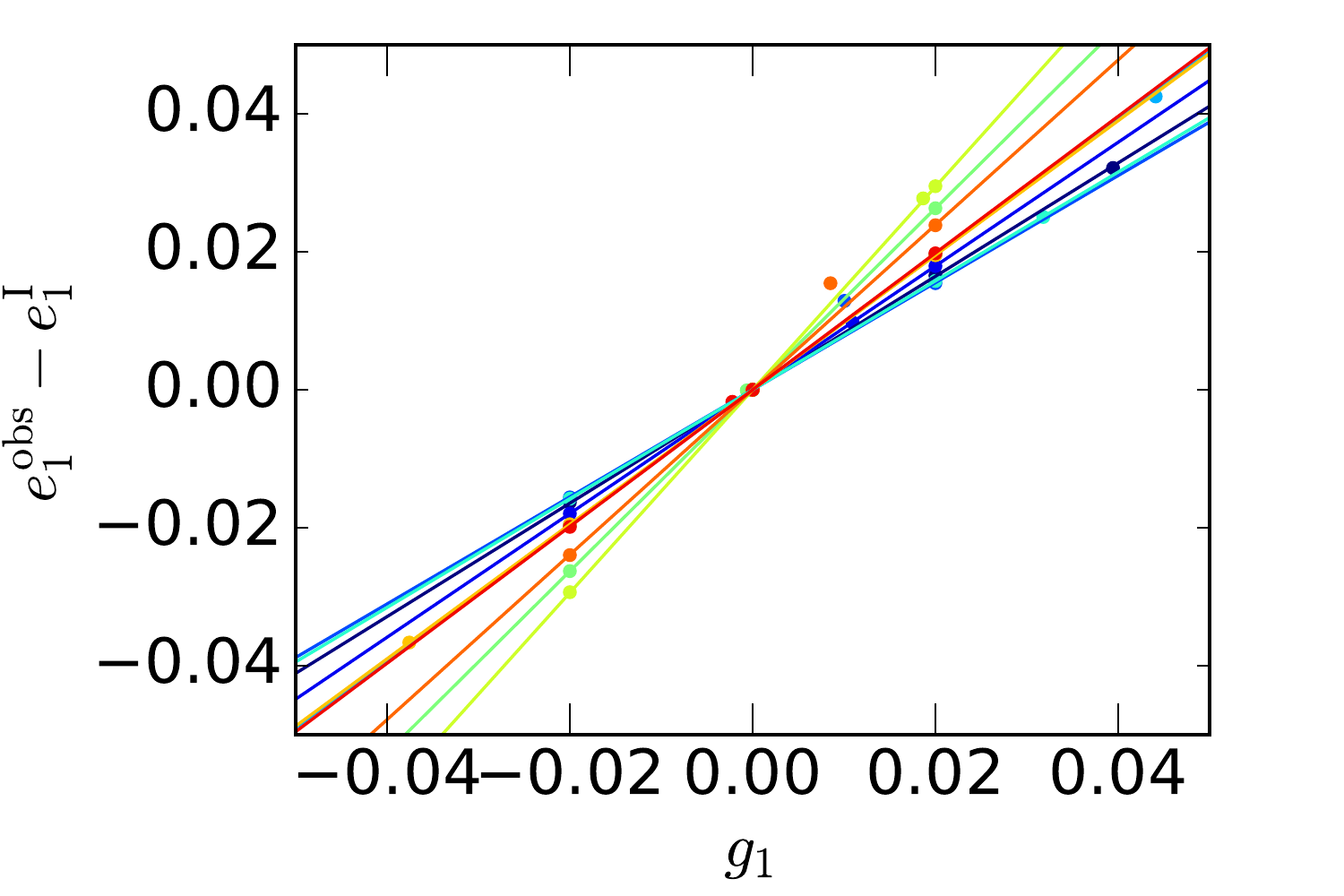}
\caption{Illustration of the shear response estimations with different shear values used for ten image examples. Each colour represents a different image, and the points show the observed ellipticities obtained with different shear values. The solid lines show $R_{11}^A$ for each case.
}
\label{fig:mcomp_fits}
\end{figure}

If the non-diagonal terms of $\mat R$ are non-zero, $R_{11}^A$ and $R_{11}^B$ should behave differently from the rest. In Fig. \ref{fig:mcomp_fits} we show ten cases of $\mat R$ estimations, where each case is represented with a different colour. The solid lines correspond to the fits of $R_{11}^A$ (so they connect the points at $g_1 = -0.02$ with those at $g_1 = 0.02$). We can see that all the points, even the random ones, tend to be well adjusted to the fitting line, although not always. These cases are an indication of non-diagonal terms of the shear response $\mat R$, causing changes in $e^{\rm obs}_1$ when changing $g_2$. In  this section we focus on the first component of shear and response, but the same holds for the second.

\begin{figure}
\centering
\includegraphics[width=.98\linewidth]{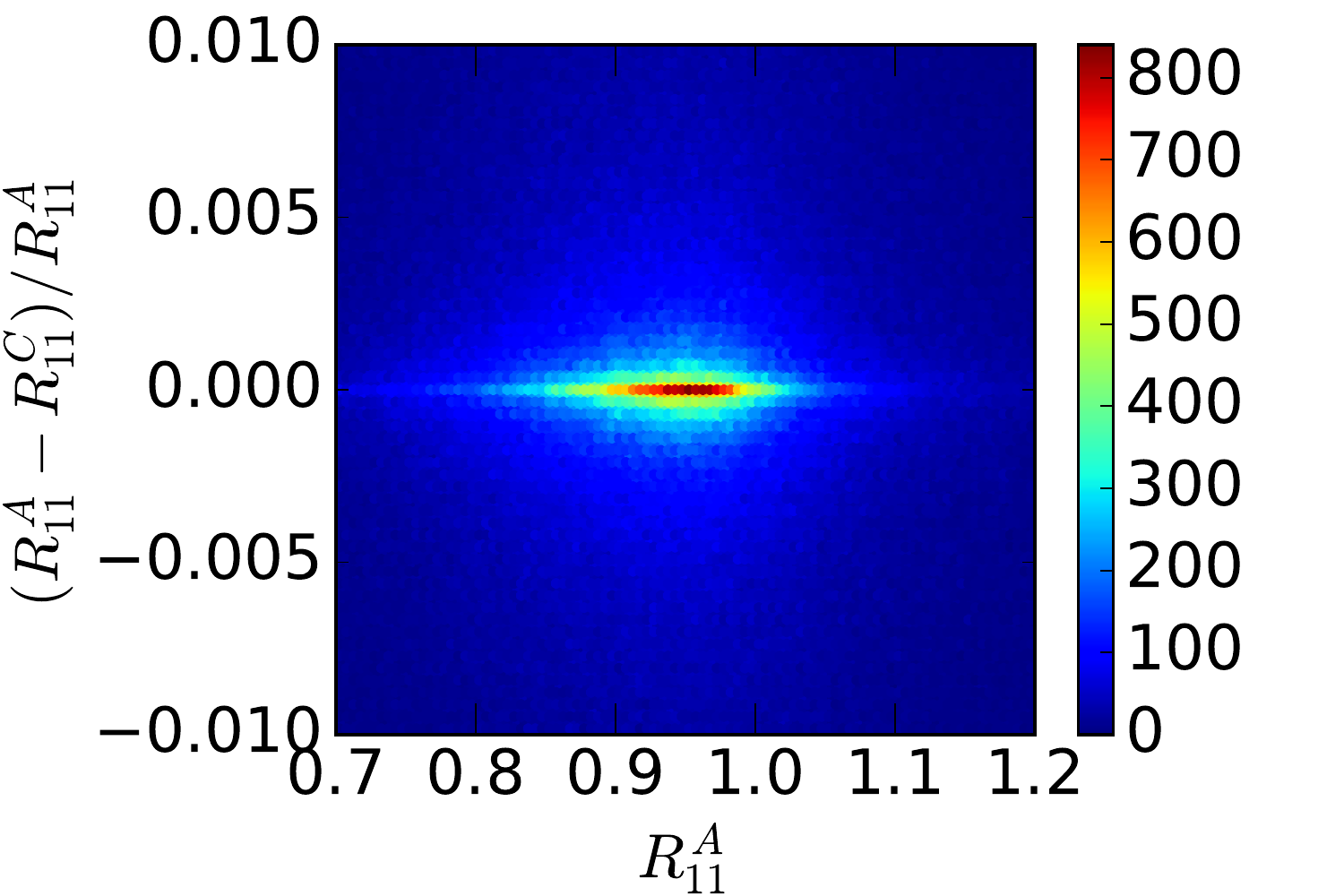}
\caption{Distribution of the differences obtained for $R_{11}^A$ and $R_{11}^C$ for our simulated images. Similar results are found for $R_{22}$.}
\label{fig:m1A_m1C}
\end{figure}

\begin{figure}
\centering
\includegraphics[width=.98\linewidth]{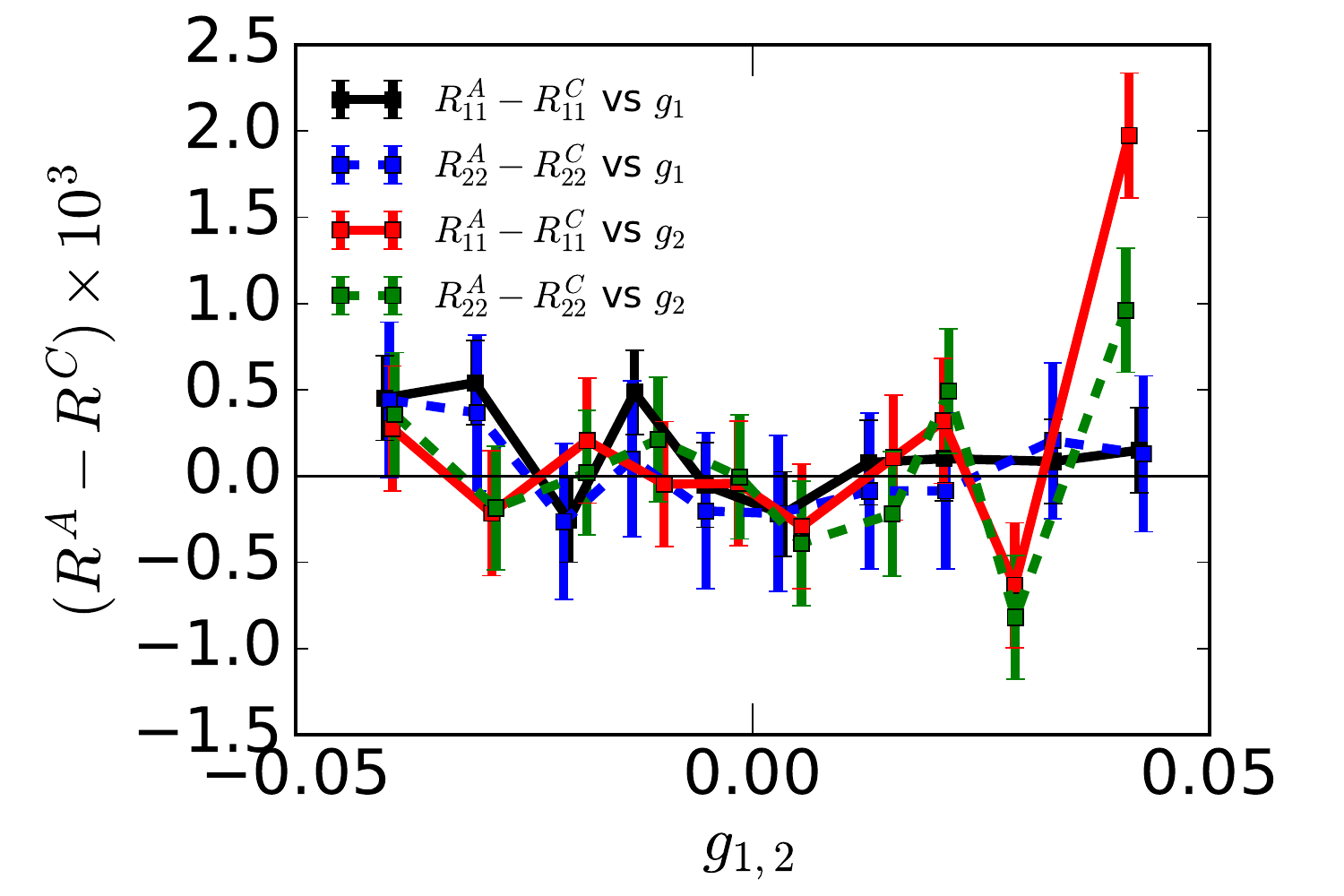}
\includegraphics[width=.98\linewidth]{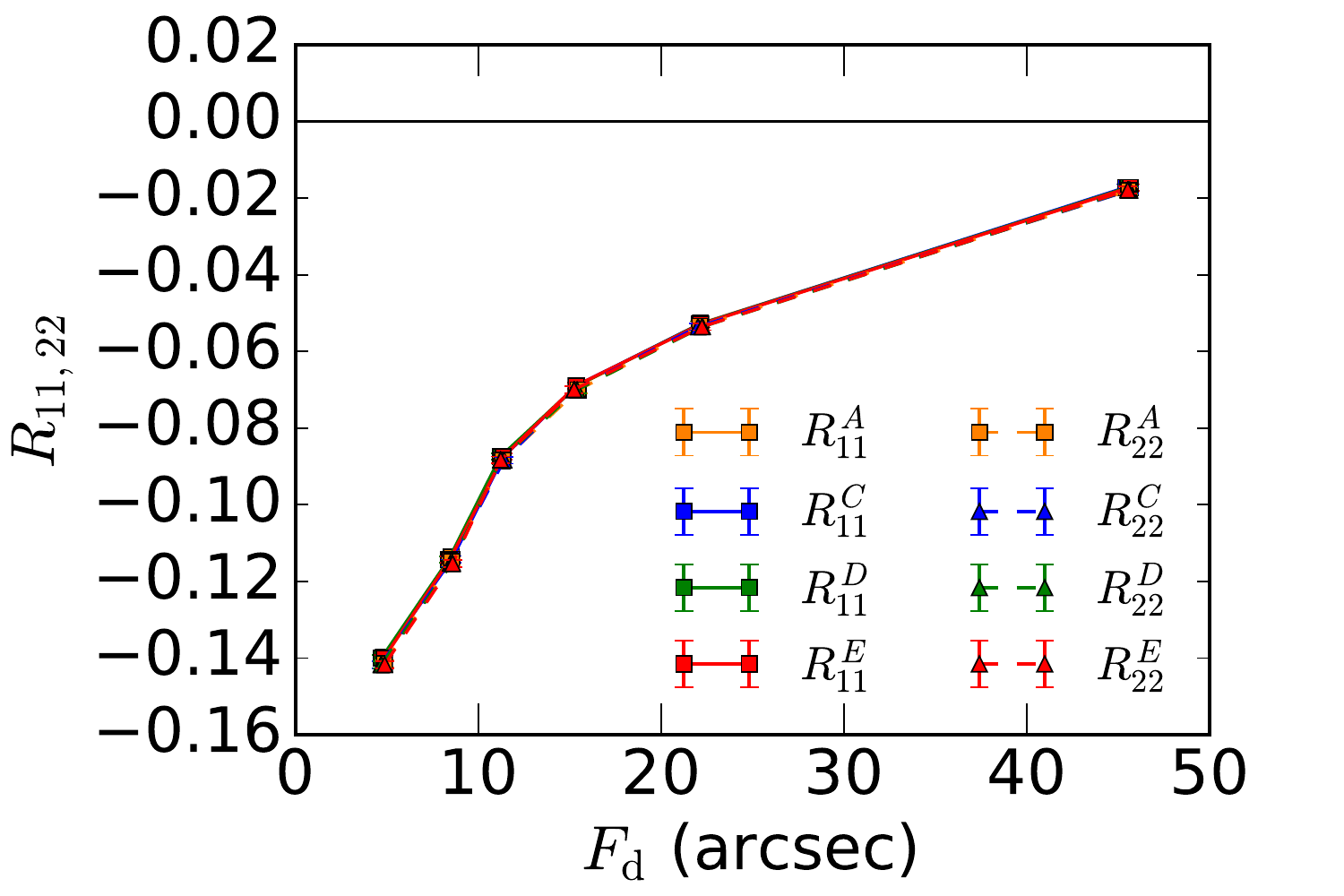}
\caption{\emph{Top panel}: Differences between $R_{11}^A$ and $R_{11}^C$ as a function of the random $g_1$ applied. \emph{Bottom panel}: Resulting average responses from the different estimations as a function of the disk flux for galaxies with a bulge and a disk. Similar results are found for $R_{22}$. }
\label{fig:m1ACDE_comp}
\end{figure}

We have found that the differences between $R_{11}^A$ and $R_{11}^B$ are negligible. This means that the method is very precise, and the relation is very well described with a linear relation. There is no need to have more than two shear values to estimate $\mat R$ precisely. When the second component is non-zero, sometimes it can affect the ellipticity measurement, in which case $R_{11}^{C,D,E}$ is different from $R_{11}^{A,B}
$.  In Fig. \ref{fig:m1A_m1C} we see the differences between $R_{11}^A$ and $R_{11}^C$. The differences between $R_{11}^D$ and $R_{11}^E$ are very similar. We see that in most of the cases the differences are negligible, uncorrelated with $R_{11}^A$ , and they average out because of symmetry. We have found that, when the second component of the shear affects $\bm e^{\rm out}$, it does it in a symmetric way so that positive $g_2$ give opposite effects to negative $g_2$. For a random distribution of $g_2$, the differences cancel out. In the top panel of Fig. \ref{fig:m1ACDE_comp} we show that the differences between the different estimations are consistent with $0$ and independent of the random $\bm g$ applied. In the bottom panel we show that the mean response as a function of the disk flux of the galaxies is consistent for all the estimators. This is actually the case as a function of all the properties studied, which means that the method is very precise and that the non-diagonal terms of the shear response do not affect the shear response estimation. As a consequence, our method does not depend on the different shear values used for the fit  as far as the shear values used are symmetric or homogeneous.

%In principle our method also suffers from ellipticity bias, since the observed
%ellipticities of the different sheared images are not equal. They are however
%very close, unlike in the case of the orthogonal pairs. We thus expect this
%ellipticity bias to be very small and negligible compared to the $2$ percent
%maximum ellipticity bias for the orthogonal pairs.

\end{document}